\documentclass[onecolumn]{article}

\usepackage{graphicx}
\usepackage{subfig}
\usepackage{algorithm, algorithmicx, algpseudocode}
\usepackage{amsmath}
\usepackage{amsthm}
\usepackage{amssymb}
\usepackage{color}
\usepackage{url}

\newcommand{\dw}{\textsc{DropWat}}

\date{}
\title{\dw{}: an Invisible Network Flow Watermark for Data Exfiltration Traceback}

\author{Alfonso~Iacovazzi, Sanat~Sarda, Daniel~Frassinelli,
        and~Yuval~Elovici% <-this stops a space
\thanks{Authors are with Singapore University of Technology and Design (SUTD), Singapore.
(e-mail: alfonso\_iacovazzi@sutd.edu.sg; sanat\_sarda@sutd.edu.sg; daniel\_frassinelli@sutd.edu.sg; yuval\_elovici@sutd.edu.sg).}}
\begin{document}

\maketitle

\begin{abstract}
Watermarking techniques have been proposed during the last 10 years as an approach to trace network flows for intrusion detection purposes. These techniques aim to impress a hidden signature on a traffic flow. A central property of network flow watermarking is invisibility, i.e., the ability to go unidentified by an unauthorized third party. Although widely sought after, the development of an invisible watermark is a challenging task that has not yet been accomplished.

In this paper we take a step forward in addressing the invisibility problem with \dw{}, an active network flow watermarking technique developed for tracing Internet flows directed to the staging server that is the final destination in a data exfiltration attack, even in the presence of several intermediate stepping stones or an anonymous network. \dw{} is a timing-based technique that indirectly modifies interpacket delays by exploiting network reaction to packet loss. We empirically demonstrate that the watermark embedded by means of \dw{} is invisible to a third party observing the watermarked traffic. We also validate \dw{} and analyze its performance in a controlled experimental framework involving the execution of a series of experiments on the Internet, using Web proxy servers as stepping stones executed on several instances in Amazon Web Services, as well as the TOR anonymous network in the place of the stepping stones. Our results show that the detection algorithm is able to identify an embedded watermark achieving over 95\% accuracy while being invisible. 
\end{abstract} 
\section{Introduction}
\label{sec:intro}
Advanced persistent threats (APTs) have received an increasing amount of attention from authorities and companies in recent years. APTs refer primarily to the high-risk threats associated with unauthorized access to a network, with the primary aim of stealing highly sensitive and valuable information. Behind every APT there usually is an adversary with specific objectives that fall into the following categories: political~\cite{li2011evidence}, economic~\cite{binde2011assessing}, technical~\cite{lee2013clustering}, and military~\cite{deweese2009capability} purposes. Although APTs are difficult to generalize, because each attack is focused on a specific target and designed accordingly, the process of implementing an APT can be broken down into six main stages which have been well described by Giura and Wang~\cite{giura2012context}: reconnaissance, delivery, exploitation, operation, data collection, and data exfiltration.
Each step in this process merits specific attention; however, in this paper we focus on the data exfiltration stage.

Data exfiltration is the last stage of an APT, and its achievement represents a successful conclusion to the entire attack process. The term data exfiltration refers to the physical process aimed at transferring previously collected sensitive data from a private device/network to an external staging server under the control of an adversary. Data exfiltration has been widely investigated~\cite{giani2006data,bertino2011towards}, and much attention has been focused on developing solutions that may prevent data exfiltration, detect a data exfiltration attack, and even nip it in the bud, before data has been stolen~\cite{liu2009sidd}. In contrast, the research community has put less effort into developing technical solutions for attack attribution, i.e., aimed at real-time identification of the adversary (individual or machine) that is attempting to obtain valuable data.

Increasingly, the process of data exfiltration is taking place over the Internet by means of digital communication between a device containing the sensitive data and the remote staging server. The adversary managing this data transfer often forwards the communication over a chain of proxy servers or an anonymous network; this is done in order to prevent others from tracing the devices under control of the adversary (by reading the destination addresses) back to the adversary, particularly when traffic flow interception has occurred.

Identifying the final destination of a data flow is a difficult problem, which is often referred to in the literature as the ``network traceback problem''~\cite{buchholz2002providing}. Network flow watermarking is a promising solution that has provided interesting insights during the last few years. Typically, watermarking solutions aim to actively modify traffic features so that they can be easily identified by a detection system, even when several noisy network nodes are crossed. Although much progress has been made in this area, two important issues remain unresolved: robustness and invisibility. Robustness refers to the property of the watermark's resistance to active noise added by an attacker to alter the watermark carrier features. Invisibility is the property of the watermark to go undetected by the adversary. Invisibility is critical, because any kind of traffic feature manipulation has potential to be easily identified by a third party (using traffic analysis instruments).

In this paper we propose \dw{}, an invisible network flow watermarking technique for data exfiltration attacks, enabling the identification of the staging server that receives the exfiltrated data. \dw{} is based on a completely new paradigm of injecting a watermark into the flow. The basic idea of our algorithm is to drop a few selected packets of a flow in order to alter the interpacket delay. We show that: 1) packet drop events can be identified, even in the presence of several stepping stones, and that they can be used as a way to convoy a watermark into traffic flows, 2) natural packet loss and intentional packet drop events in the network cannot be distinguished from each other, and 3) the watermark embedded with our algorithm is invisible. We evaluate \dw{} under different network scenarios with different conditions of packet loss and throughput, on real traffic on the Internet.

The rest of the paper is organized as follows. Section~\ref{sec:rel_work} provides an overview of previous work on network flow watermarking and its application in overcoming the traceback problem. The attack scenario and reference architecture are described in Section~\ref{sec:scenario}. The \dw{} embedding and detection algorithms are described in Section~\ref{sec:dropwat}. Section~\ref{sec:invisibility} contains an in depth discussion and analysis of the invisibility property. Section~\ref{sec:performance} provides a description of our experimental results and validation of the effectiveness of \dw{}. In Section~\ref{sec:discussion} we discuss some critical aspects of our watermarking algorithm, and our conclusions are in Section~\ref{sec:conclusion}. 
\section{Related work}
\label{sec:rel_work}
The traceback problem, aimed at identifying the real destination of a traffic flow, has been extensively investigated~\cite{savage2000practical,song2001advanced,sung2003ip,li2004large,choi2003network,mitropoulos2005network,hamadeh2006taxonomy}. In 2001, Wang et al. introduced network flow watermarking as a possible means of overcoming the traceback problem~\cite{wang2001sleepy}. Since then, many network flow watermarking algorithms have been developed and proposed. Recently, Mazurczyk et al.~\cite{mazurczyk2016information} and Iacovazzi et al.~\cite{iacovazzi2017network} presented surveys providing a comprehensive analysis and comparison of the main network flow watermarking solutions known in the literature.

The vast majority of the proposed techniques modify the packet timestamps in order to impress a specific timing pattern onto the network flows~\cite{wang2003robust1,peng2005active,pyun2007tracing,houmansadr2009rainbow,wang2007network,gong2012invisible,houmansadr2011swirl,houmansadr2013botmosaic}. RAINBOW is an example of a timing-based watermarking algorithm~\cite{houmansadr2009rainbow}, where each packet is delayed by a computed value; the delay values equal the output of a cumulative function which randomly evolves with a step of plus/minus a specified watermark amplitude per each packet. RAINBOW's detection algorithm is based on the comparison between the interpacket delays (IPDs) of the flow before being watermarked and those of the flows intercepted by the detector.

The technique proposed by Peng et al.~\cite{peng2005active} is also based on IPDs. The authors consider two groups of randomly selected pairs of consecutive packets; the IPDs are computed for every pair in each group. The two average values of IPDs in the two groups are considered statistically equal to each other. Their proposed watermarking algorithm aims to slightly modify the IPDs, so that the difference between the two average values is not zero. The numerosity of the two groups represents a kind of redundancy and determines the detection reliability.

A technique called interval centroid-based watermarking was introduced by Wang in 2007~\cite{wang2007network}. In this technique, the time axis is divided within intervals of fixed duration $T$. A centroid is computed for each interval as the average value of the remainders remaining after dividing the timestamps of packets observed in that interval by $T$. In the embedding algorithm, some packets of the flow are delayed so that the statistical balances among groups of intervals are altered. Watermark detection is based on the statistical analysis of interval centroids. A variety of similar methods have been also suggested by other researchers~\cite{houmansadr2011swirl,luo2012interval,wang2010double}.

In interval packet counting-based techniques, the time axis is divided within intervals~\cite{houmansadr2013botmosaic,pyun2007tracing,zand2014rippler}; the number of packets in each interval is the carrier of the watermark, and some packets of the flow are delayed in order to alter the statistical balance of the packet counting per interval.

Timing-based algorithms are very attractive, because packet timing can easily be modified by the watermarker without having to access the data at any protocol level. Nevertheless, timing can also be altered by natural network perturbation or be artificially modified by an attacker, resulting in the failure of watermark detection. For this reason, other watermarking algorithms have been created that are robust against timing perturbation~\cite{wang2003robust1,peng2005active}, repacketization~\cite{pyun2007tracing}, and chaff packet injection attacks~\cite{peng2005active,houmansadr2009rainbow}.

One major drawback of timing-based schemes is that they primarily target flows with less than 50 packets per second (PPS). If, for example, we consider a scenario such as an illegal data transfer in which the transfer rate can easily be 200 -- 500 PPS or more (assuming 1500 bytes/packet, and a speed of 300 -- 750~KB/s), these algorithms would not be effective. The reason for this is that most of the parameters have to be re-adapted in order to cope with the higher network speed and lower IPD. However, at higher network speed, proxy servers tend to obfuscate any kind of slight timing perturbation, making small changes impossible to detect. One could argue for the use of more significant perturbations, but this would make the watermark more visible and significantly impact the performance of the network (and not necessarily improve the detection rate, since generally the parameters need to be chosen proportionally to the IPD).
The only technique that would be able to work with bulk traffic is the centroid-based solution developed by Wang~\cite{wang2007network}, but it would also require a lot of buffering and TCP level multi-flow analysis which makes it impracticable to implement in border routers where network speed, memory, and computational power are generally strict constraints.

Timing is not the only feature that can be used as a watermark carrier; packet size~\cite{ramsbrock2008first,ling2013novel,arp2015torben} and bit rate~\cite{chan2013revisiting,yu2007dsss} are two traffic features that have attracted attention as well. However, size-based watermarks need to be embedded directly at the source of the traffic flow, while rate-based watermarks are strongly visible to third parties. 

Invisibility (the capability of passing unnoticed by an attacker) is one of the most important properties of a watermark algorithm. Although some researchers have designed watermarking algorithms that were claimed to be invisible~\cite{yu2007dsss,houmansadr2009rainbow,gong2013invisible}, later studies have empirically shown that a completely invisible watermark does not exist yet~\cite{kiyavash2008multi,luo2010secrecy,luo2011exposing,lin2012new,jia2013blind}.

\section{Attack scenario}
\label{sec:scenario}
\begin{figure*}
\begin{center}
  \includegraphics[width=0.6\textwidth]{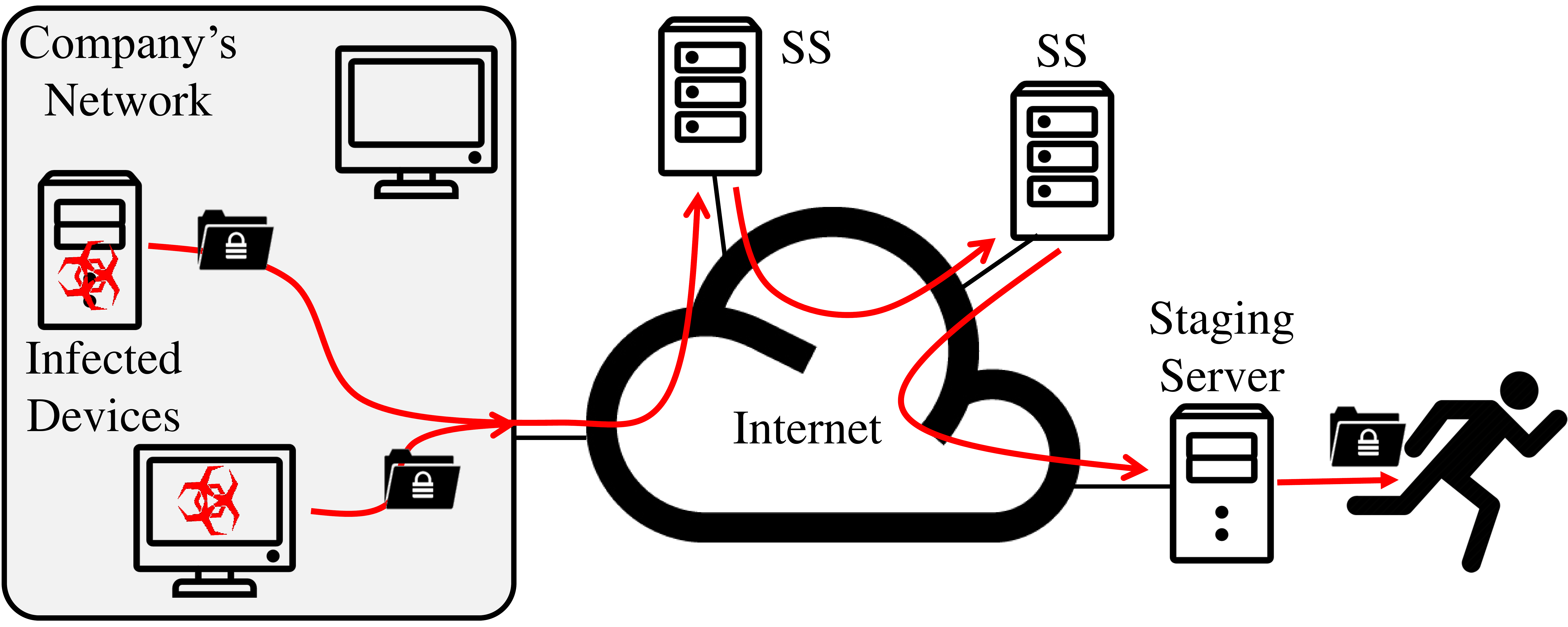}
  \caption{Attack scenario.}\label{fig:attack-scenario}
\end{center}
\end{figure*}

\subsection{Data exfiltration attack}
We consider the scenario shown in Figure~\ref{fig:attack-scenario} in which an adversary wants to take possession of confidential data, files, or documents that belong to a person or company and are stored in digital format on a device connected to the Internet in some way. These documents can be sensitive, private, copyrighted, or accessible only with required permission. In our scenario, the attacker has managed to install a malware on the targeted device. This malware allows the attacker to control the device and exfiltrate data from the private network to an external server (staging server) under her control, via an Internet connection. Two or more stepping stones are used in order to disallow possible identification of the staging server (its IP address, IP address geolocation, etc.). Once the targeted data is saved on the staging server, the attacker is able to access the data at any time. If the staging server is identified, the attacker may be identified as well, when it connects to the server.

\subsection{Stepping stones}
\label{sec:stepping_stone}
A stepping stone (SS), also referred to as a proxy server, is an intermediary device or application interposed in the communication between two hosts in a network. The main purpose of an SS is to prevent the identification of the real sender and/or recipient of the exchanged messages, in the event that a third party intercepts the communication. The property of a flow to not be associated with the communication's real endpoints is known as the ``unlinkability'' of the sender and receiver. In this case, whenever a client wishes to contact a server for Web content, it does not send messages directly to the server, but instead it connects and sends the messages to a proxy server which is responsible for forwarding the traffic to the real recipient. Conversely, reply messages from the server to the client will first be delivered to the SS and then be forwarded to the client. In most cases, communications to and from an SS are based on encrypted and authenticated connections. Thus, the integrity of unlinkability property is preserved when a third party observes the traffic in the middle of one of the two connections involved; nevertheless, the communication is vulnerable to passive attacks performed on the proxy server. A single point of vulnerability can be avoided by using two or more SSs in a chain.

\subsubsection{Implementation and packet loss propagation}
There are many types of SSs and ways of implementing them: Web proxy servers, TOR software, etc.~\cite{edman2009anonymity}. An explanation of different SS implementations and a description of their operations are not within the scope of this paper; we prefer to focus on how the implementation of a SS may influence traffic patterns in cases in which a packet loss occurs before reaching the SS. In these cases, the SS can behave as the propagator or retriever of lost packets. The SS behavior depends on the combination of two factors: 1) the protocols used for transferring the traffic, and 2) the protocol layer at which the SS operates. For example, let us consider communication over TCP: when the SS handles data units at the transport layer, two independent TCP connections are established, one from the client to the SS, and the other from the SS to the server; when a packet directed to the SS is lost, the SS notices that a packet is missing and requests retransmission, so the loss is not propagated. Thus, here the SS acts as a retriever. Alternatively, an SS can also be implemented to work at the network layer (such as an NAT service). In this case, the source and destination of transport layer segments retain the real communication's source and destination. Here the SS is only responsible for being an intermediary at the network layer. The two endpoints send their IP packets to the SS which decapsulates transport segments from packets, makes port translation, and encapsulates each segment in a new IP packet containing the SS's IP address in the source address field and the real destination's IP address (or the next hop's IP address in case of a chain of SSs) in the destination address field. Here the SS changes the transport layer ports, but it does not interfere with the operations performed by the transport protocol which means that packet loss is propagated to the next hop of the path. Thus, in this scenario, the SS acts merely as a propagator.

In this paper we refer to an attack scenario in which SSs do not propagate packet loss, as this scenario is used by most attackers by implementing their own proxy networks or using TOR because it does not leave a trace of real IP of their staging server. Nevertheless, a slightly modified version of our algorithm would work in cases of SSs propagating loss.

Without loss of generality, hereafter we base our analysis on a scenario in which communications travel over TCP, and the SS operates at the transport layer.

\section{\dw{} algorithm}
\label{sec:dropwat}
In this section we describe \dw{}, a watermarking technique based on packet dropping, which indirectly modifies IPDs of selected packets. The basic idea of our technique is to mimic a natural network behavior, namely packet loss events caused by a single bottleneck node, and exploit it as a watermark identifiable despite the traffic flows crossing one or more SSs.

An attacker is not able to distinguish between naturally lost packets and those intentionally dropped, because both events cause the same behavior in a network.\footnote{We use the term \emph{intentionally dropped packets} to indicate only those packets that are dropped in order to embed a watermark in the traffic flow. Packets dropped  due to other causes (e.g. buffer overflow, framing error, etc.) are considered \emph{naturally lost packets}.} If the attacker is unable to distinguish between a sequence of lost packet events due to a real bottleneck node and a sequence of dropped packet events caused by an emulated bottleneck node, then the watermark will be invisible.

In the following subsections we explain what happens in our scenario when a packet loss event occurs; we then provide a detailed description of \dw{}, our proposed watermarking method for tracing data exfiltration attacks.

\subsection{Packet loss occurrence}
\begin{figure}
\begin{center}
  \includegraphics[width=0.48\textwidth]{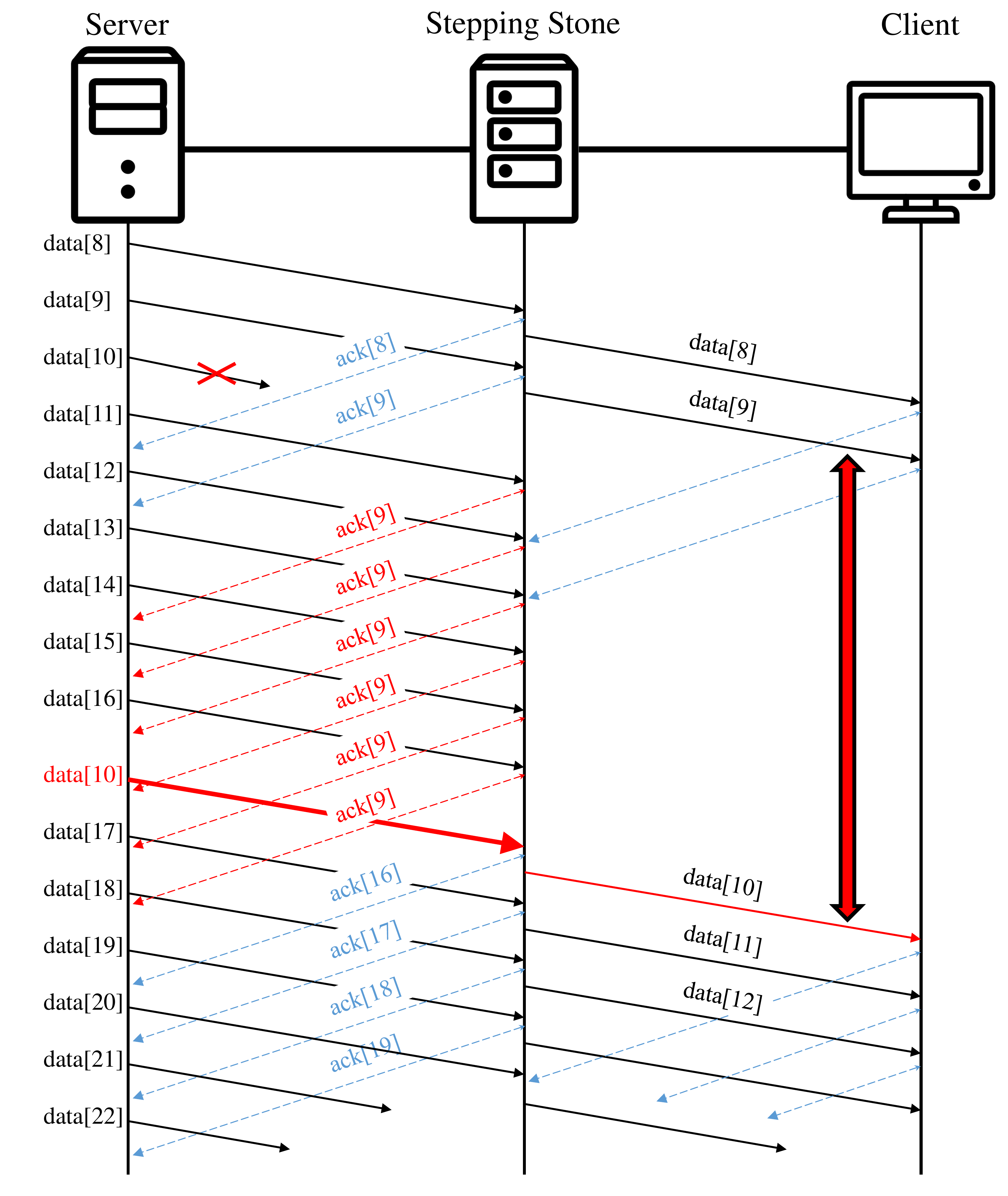}
  \caption{Packet loss event in a scenario with one SS.}\label{fig:tcp-loss}
\end{center}
\end{figure}

Packet losses occur naturally in computer networks and are caused by several reasons, such as faulty hardware or cabling, buffer overflow due to link or node congestion, data corruption due to components with high bit error rates, packet filtering, etc. Internet protocol (IP) provides a service of best effort delivery; it does not deal with detecting and recovering lost packets. The management of packet recovery for reliable delivery is left to higher layer protocols. Recovery of lost packets can be guaranteed at the transport layer with the TCP protocol. 

The behavior of an SS handling data units at the transport layer in a case of packet loss is depicted in Figure~\ref{fig:tcp-loss}. The left half of the figure shows the typical TCP behavior when a packet is lost. It can be seen that the SS sends duplicated acknowledgements until it receives the expected packet. The server keeps sending subsequent packets until it realizes that a loss has occurred, and then it re-sends the lost packet. The time required before re-sending a lost packet depends on the TCP implementation used by the sender. When fast retransmission is adopted, a packet is sent a second time after receiving a specified number of repeated acknowledgements (usually set to three in the most commonly used TCP stack implementations). Since the TCP connection endpoint is at the SS, the TCP protocol reorganizes out of order data at the SS application layer, so that $data[11]-data[16]$ cannot be delivered until $data[10]$ is correctly received. For this reason, when a packet is lost, the SS cannot keep sending data to the next hop even though out of order packets are received by the TCP protocol. This entails that the IPD between $data[10]$ and $data[9]$ at the destination will be altered and equal to a value greater than the round trip time from the server to the SS.

\begin{figure}
\begin{center}
  \includegraphics[width=0.32\textwidth]{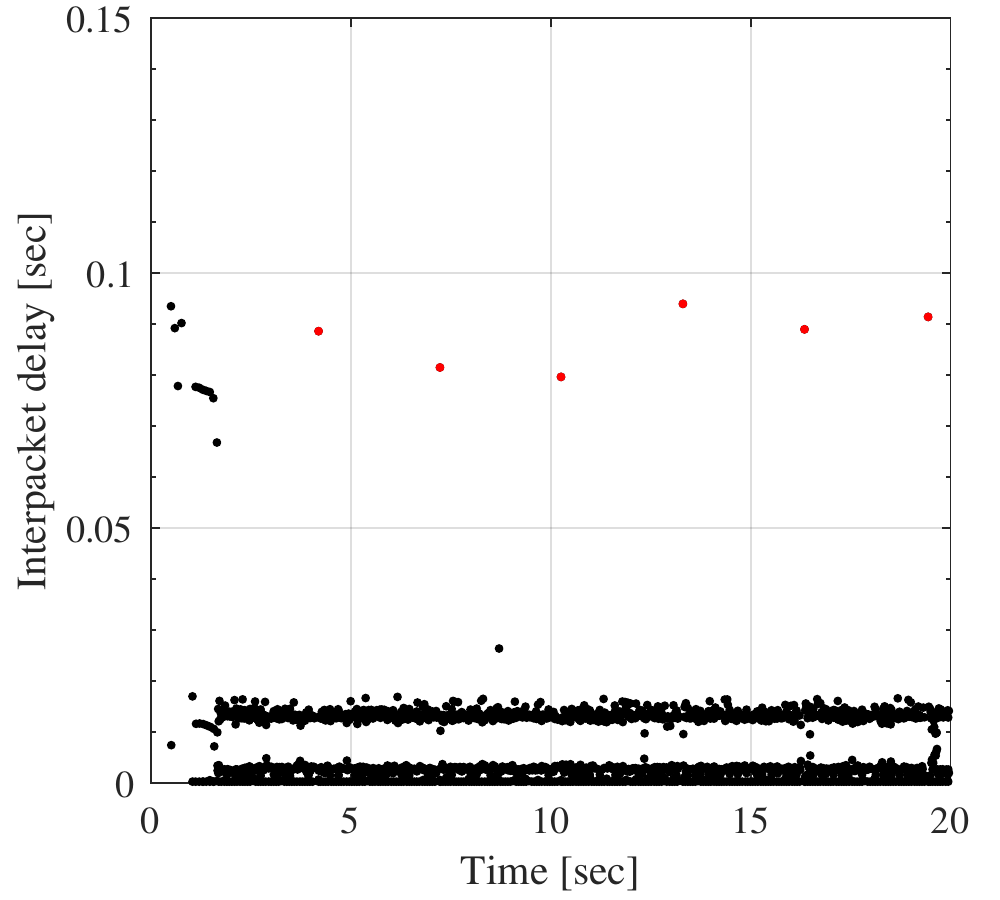}
  \caption{The impact of packet loss events on IPDs measured on the client side.}\label{fig:ss-loss}
\end{center}
\end{figure}

In Figure~\ref{fig:ss-loss} we show the trend of the IPDs measured at the client endpoint, when a 50~MB file is downloaded from the server. The communication is intermediated by two SSs. A packet was periodically dropped in the connection between the server and the first SS encountered. The round trip time (RTT) between the two was 80~ms. During the first few seconds of the communication, IPDs are affected by the TCP's slow start. After the slow start phase, the system reaches a stable state in which the IPDs maintain regular values. The regularity is broken when a packet loss event occurs, as highlighted in the figure. The trend is maintained even in the presence of multiple SSs. Thus, we can claim that although the packets are sent sequentially from the SS, any packet lost (and later retrieved) in the first connection can be identified in the second connection by analyzing IPDs on the client side.

\begin{figure}
\begin{center}
  \includegraphics[width=0.32\textwidth]{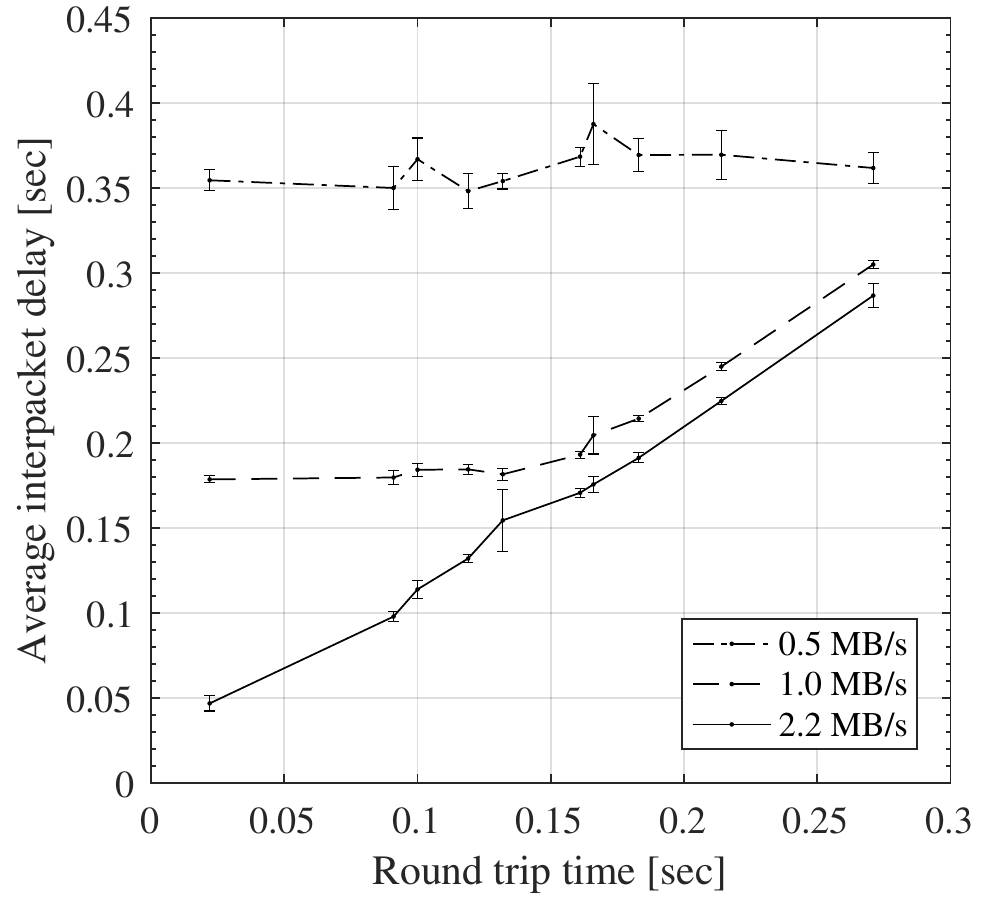}
  \caption{Average value of the IPDs that correspond to the packet loss events plotted as a function of the RTT from the server to the SS.}\label{fig:avg-delay}
\end{center}
\end{figure}

The server packet transfer rate and the RTT between the server and the first SS may change the effect of packet loss events on IPDs. To give an idea of this effect, we averaged the
values of the IPDs that correspond to the packet loss events measured on the client side, and we plotted them in Figure~\ref{fig:avg-delay} as a function of the RTT from the server to the SS. The graph shows a linear trend when the transfer rate $R$ is 2.2~MB/s, a constant and later linear ramp for $R=$ 1~MB/s, and a constant trend for $R=$ 0.5~MB/s. This is due to the fact that IPDs, altered by packet loss, are a function of both transfer rate and RTT.

\begin{figure}
\begin{center}
  \includegraphics[width=0.32\textwidth]{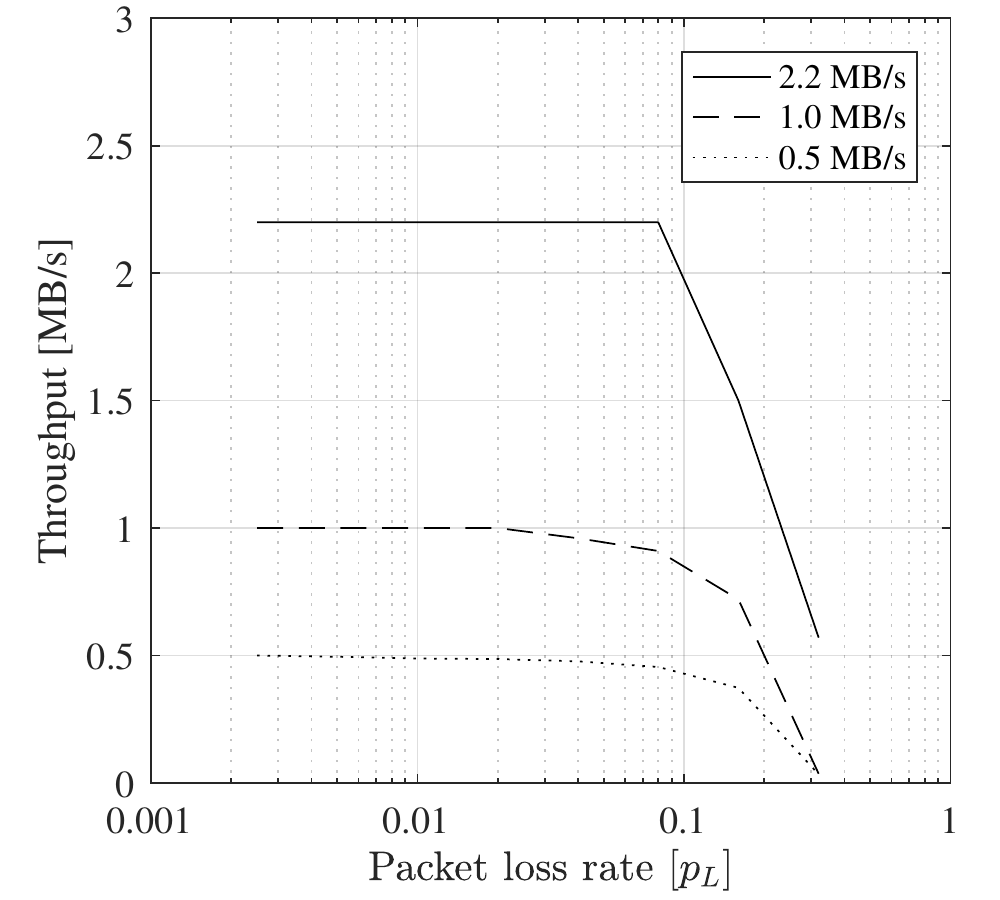}
  \caption{Throughput as a function of packet loss rate.}\label{fig:bw}
\end{center}
\end{figure}

When the number of lost packets in the network becomes high, the TCP protocol interprets this behavior as network congestion and reacts by reducing the rate at which packets are sent. The reduction of the throughput caused by varying the packet loss rate is shown in Figure~\ref{fig:bw}.  Thus, in order to ensure that the embedded watermark does not have a significant impact on the network performance, the packet loss rate should be less than 1\%. 

\subsection{Watermarking architecture}
\begin{figure}
\begin{center}
  \includegraphics[width=0.48\textwidth]{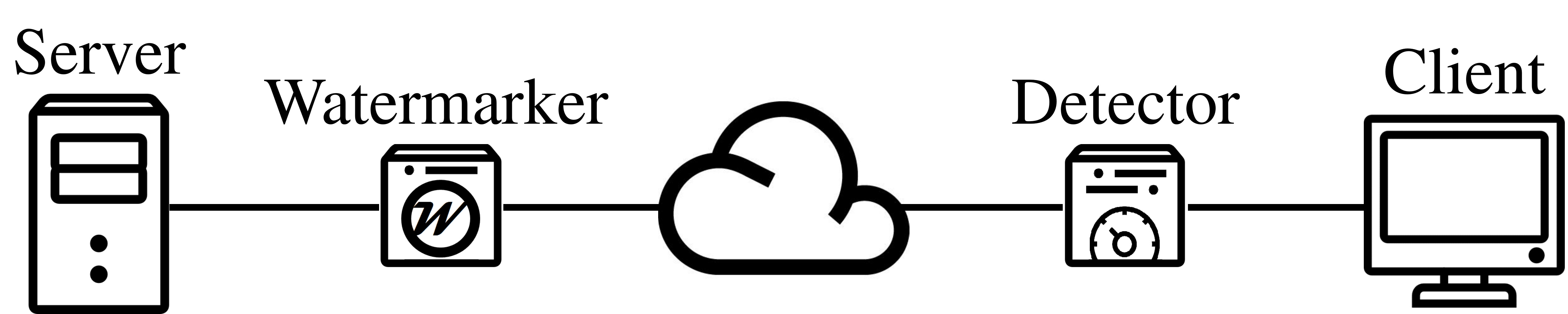}
  \caption{Architecture of the \dw{} watermarking system.}\label{fig:wm-arch}
\end{center}
\end{figure}

The architecture of \dw{} is similar to other existing active network flow watermarking techniques. As shown in Figure~\ref{fig:wm-arch}, the system is composed of a watermarker and a detector. The watermarker intercepts targeted flows and embeds the watermark. In our case, this action corresponds to selectively dropping some packets. The detector observes and analyzes traffic flows, and looks for the presence of a watermark. In the following two subsections the watermark embedding and detection algorithms are described in greater detail.

\subsection{Watermark embedding}
\label{sec:embedding}
\dw{}'s embedding algorithm aims at dropping pseudo-randomly selected packets so that the sequence of dropped packets looks like a loss sequence caused by a single bottleneck node.
A single bottleneck node can be described as a buffer which can hold a specific number of packets. An input process fills the buffer with  packets coming from several sources; an output process extracts packets from the buffer at a fixed rate limited by the output link rate. When an incoming packet finds the buffer full, it will be discarded and a loss event will occur.

\begin{figure*}
\begin{center}
  \includegraphics[width=0.80\textwidth]{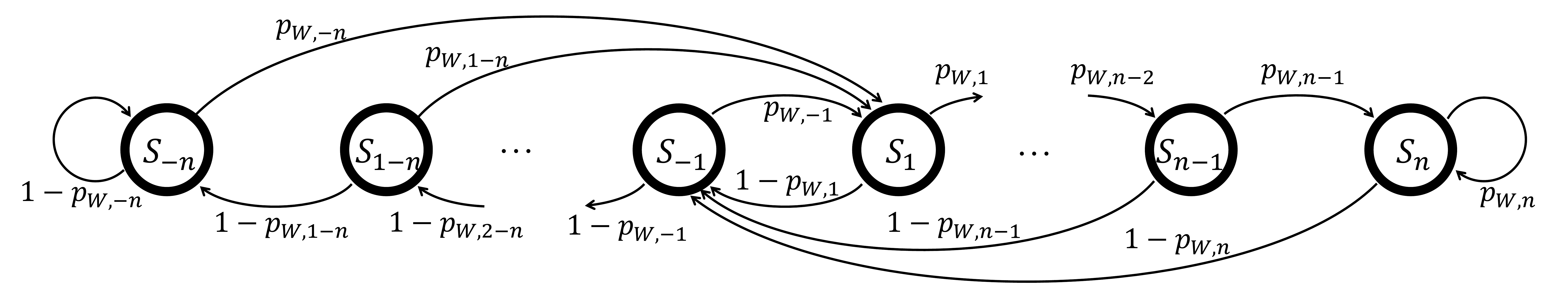}
  \caption{Variation on the extended Gilbert model.}\label{fig:gilbert}
\end{center}
\end{figure*}

In order to emulate the behavior of a single bottleneck node, we model packet loss behavior according to a modified version of the extended Gilbert model. The extended Gilbert model was used to reflect packet loss behaviors in noisy networks by Sanneck et al.~\cite{sanneck1999framework}, and Yu et al.~\cite{yu2005accuracy} demonstrated that this model very well approximate the packet loss behavior of a single multiplexer. Let $X_i$ be the binary event for the $i$-th packet of a flow, which can assume the value 1 for a dropped packet, and 0 for a non-dropped packet. In our modified version of the extended Gilbert model, an event state is assumed to be dependent on the last run composed of up to $n$ consecutive identical events. In this model (hereafter referred to as $\mathcal{W}$) we need only $2n$ different states, and it can be completely described by the set of probabilities $\{p_{W,-n},p_{W,1-n},\dots,p_{W,-1},p_{W,1},\dots,p_{W,n}\}$. The model of packet drop states is depicted in Figure~\ref{fig:gilbert}.

\begin{figure}
\begin{center}
  \includegraphics[width=0.4\textwidth]{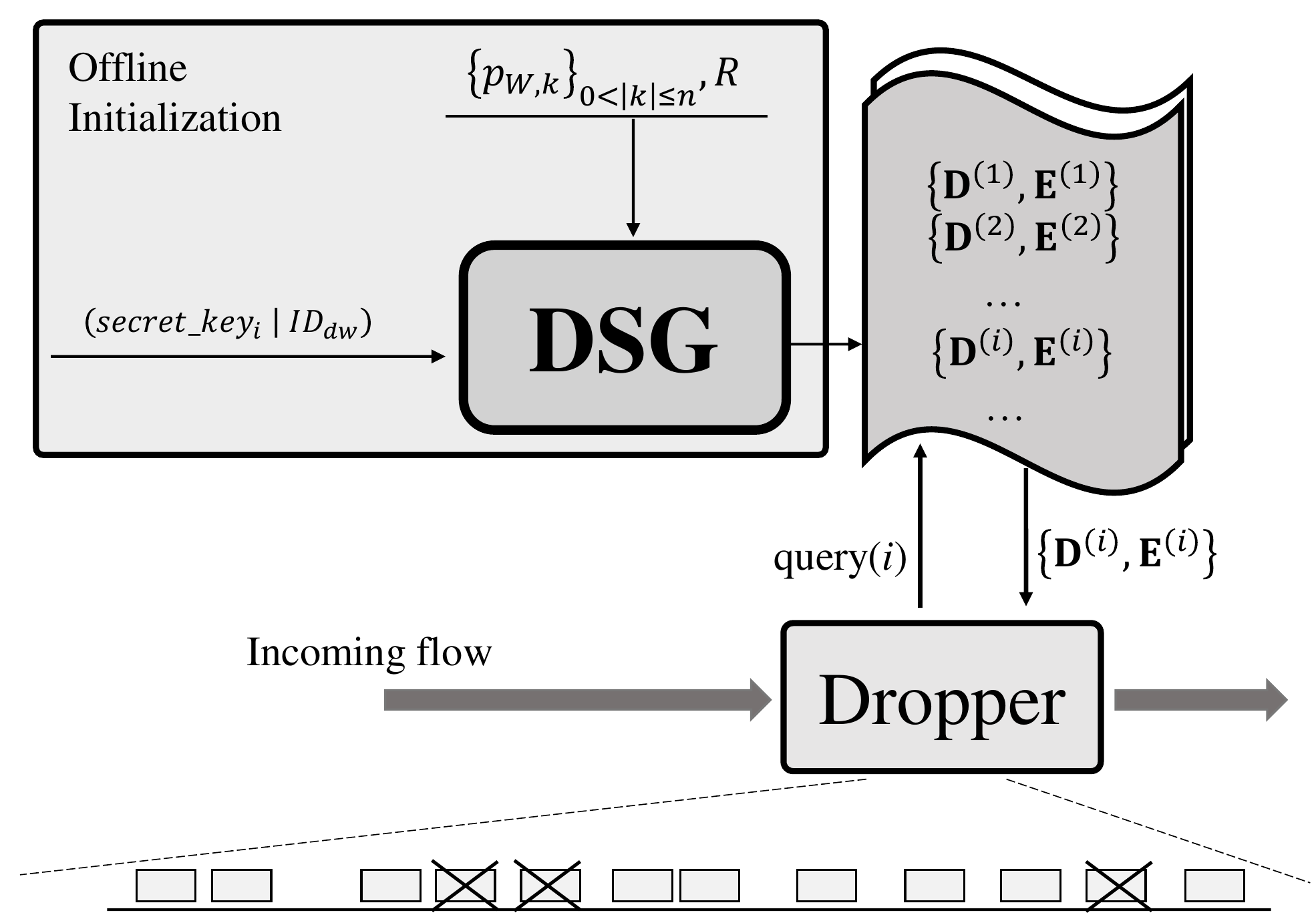}
  \caption{\dw{} embedding scheme.}\label{fig:dropper}
\end{center}
\end{figure}

The watermarking process, as depicted in Figure~\ref{fig:dropper}, can be divided into two parts: offline initialization and online packet dropping. The algorithm evolves as a periodic process with time period $T$. Let $T_0=0$ be the zero time reference, we indicate the starting time of the $i$-th time period as $T_i=i T$.

The offline initialization takes as input: 1) the model probabilities $\{p_{W,-n},$ $p_{W,1-n},\dots,p_{W,-1},p_{W,1},\dots,p_{W,n}\}$, 2) a secret key shared with the watermark detector, 3) a watermarker identifier $ID_{dw}$, and 4) the reference throughput $R$. The concatenation of the secret key and $ID_{dw}$ will be used as the seed of the dropping sequence generator (DSG), a cryptographically secure function generating a pseudo-random binary sequence (sequence of events) which follows the model $\mathcal{W}$. Let $\mathbf{B}^{(i)}=[b_1^{(i)},b_2^{(i)},\dots,b_{N}^{(i)}]$ indicate the $i$-th binary sequence generated by the DSG, where $ N=\lceil R\cdot T/ L_{ref}\rceil$ is the expected number of packets in the period, $L_{ref}$ is the reference packet size computed as the maximum transmission unit (MTU), and $\Delta t_{pkt}=L_{ref}/ R$ is the time required to send a packet. The throughput $R$ can be set at the maximum transfer rate of the watermarker.

The DSG can be efficiently implemented by using two secure pseudo-random number generators (PRNG). The first, $prng_{syn}$, is used to synchronize the watermarker and the detector, as shown in Algorithm \ref{alg:syncDSG}, and is initialized using the shared key $shared\_key = secret\_key | ID_{dw}$ and the initial time $T_0$. 
\begin{algorithm}
\caption{Synchronization}
\label{alg:syncDSG}
\begin{algorithmic}[1]
\Procedure{syncDSG}{$shared\_key$, $T_0$, $T$}
\State 	$prng_{syn} \gets 	\textbf{new} \;\; PRNG(shared\_key) $
\State  $T_{curr} 	\gets 	T_0$
\While	{$T_{curr} < system.timeNow()$}
	\State 	$prng_{syn}.genRand() $ \Comment{Generate a pseudo-random number in $[0,1)$}
	\State 	$T_{curr} \gets T_{curr} + T $
\EndWhile
\State	\textbf{return} $prng_{syn}$
\EndProcedure
\end{algorithmic}
\end{algorithm}

After every time period $T$, a new $seed = prng_{syn}$ is generated by Algorithm \ref{alg:syncDSG} and used to initialize a second $prng_{dsg}$. This newly created $prng_{dsg}$ is used to generate a valid binary sequence $\mathbf{B}^{(i)}$ of length $N$ by executing Algorithm \ref{alg:genDSG}.
\begin{algorithm}
\caption{Binary Sequence Generation}
\label{alg:genDSG}
\begin{algorithmic}[1]
\Procedure{genDSG}{$prng_{syn}$, $\{p_{W,k}\}_{0<|k|\leq n}$, $N$}
\State $seed \gets prng_{syn}.genRand()$
\State $prng_{dsg} \gets \textbf{new} \;\; PRNG(seed)$
\State $\mathbf{B} \gets \textbf{new} \;\; vector()$
\State	$k \gets -n$
\While{$\mathbf{B}.size < N$}
	\If{$ prng_{dsg}.genRand() < p_{W,k}$}
		\State $\mathbf{B}.append(1)$
        \State $k = \max\{1,\min\{k+1,n\}\}$
	\Else
		\State $\mathbf{B}.append(0)$
        \State $k = \min\{-1,\max\{k-1,-n\}\}$
	\EndIf
\EndWhile
\State	\textbf{return} $\mathbf{B}$
\EndProcedure
\end{algorithmic}
\end{algorithm}

\begin{algorithm}
\caption{Dropping sequence conversion}
\label{alg:dsc}
\begin{algorithmic}[1]
\Procedure{DSC}{$\mathbf{B},\: \Delta t_{pkt}$}
\State $\mathbf{D},\: \mathbf{E} \gets \textbf{new} \;\; vector()$
\While{$k <= \mathbf{B}.size$}
    \State $n \gets 1$
	\If{$ \mathbf{B}[k] == 1 $}
        \State $n \gets countOnes(k,\: \mathbf{B})$  \Comment{Count consecutive ones from position $k$ in $\mathbf{B}$}
		\State $D.append(k\cdot \Delta t_{pkt})$
        \State $E.append(n\cdot \Delta t_{pkt})$
    \EndIf
    \State $k \gets k + n$
\EndWhile
\State	\textbf{return} $\mathbf{D},\: \mathbf{E}$
\EndProcedure
\end{algorithmic}
\end{algorithm}

The binary sequence is then converted to a dropping sequence. A dropping sequence corresponds to a sequence of packet dropping time intervals, and it is described by two vectors $\mathbf{D}^{(i)}=[d^{(i)}_1,d^{(i)}_2,\dots,d^{(i)}_{K_i}]$ and $\mathbf{E}^{(i)}=[e^{(i)}_1,e^{(i)}_2,\dots,e^{(i)}_{K_i}]$ of length $K_i$, where $d^{(i)}_k$ and $e^{(i)}_k$ indicate the starting time and the duration, respectively, for the $k$-th dropping time interval, expressed in nanoseconds, and $K_i$ is the number of dropping intervals in the $i$-th time period. The dropping sequence conversion is performed by means of Algorithm \ref{alg:dsc}.

The dropper works by discarding all of the packets traversing the watermarker during any dropping time interval. All of the other packets will be correctly forwarded to the proper interface.

\subsection{Watermark detection}
\label{sec:rainbow}
The detector is placed at one or more points in the network where we might expect to observe watermarked flows. It analyzes all traffic and tries to understand whether a watermark is embedded in any of the observed flows.

The detector is aware of the input data to the DSG and the cryptographical function used by the watermarker, so it can compute all of the dropping time intervals. The detector analyzes the IPDs for packets observed during the dropping time intervals, and for each flow it builds the sequence of identified lost packets. If a significant percentage of lost packets of a flow are detected during the dropping time intervals, the flow is suspected of being watermarked.\footnote{Since burst losses are managed by the TCP protocol through burst retransmissions, the detector can only identify the first dropped packet of a burst. For this reason, the burstness of packet loss is not relevant to detection.} The detector and the watermarker must be accurately synchronized in order to agree on the valid dropping sequence for a time period; to maintain synchronization over a long period of time, an external synchronization server (such as NTP) may be used to reset the internal clocks of the two devices.

The watermark detection algorithm can be summarized in three main steps: 1) IPD computation, 2) outlier detection, and 3) watermark identification. The three steps are only performed for the packets observed during the dropping time intervals.

\begin{itemize}
  \item \emph{IPD computation.} An IP flow is sniffed, and packet timestamps are measured. A nominal task, IPD computation is based on the difference of consecutive packet timestamps.
  \item \emph{Outlier detection.} IPDs are analyzed to identify packet loss events. The identification is based on a simple outlier detection algorithm. Let $\hat{t}_k$ be the timestamp of the $k$-th packet observed at the detector, $\Delta \hat{t}_k=\hat{t}_k-\hat{t}_{k-1}$ be the $k$-th IPD, and $v$ be a comparison window size. $\Delta \hat{t}_k$ is considered an outlier if $(\alpha\cdot\Delta \hat{t}_k)>\Delta \hat{t}_h$ for all $h\in\{k-v,\dots,k-1,k+1,\dots,k+v\}$, with $0<\alpha<1$. The observation times of the outlier packets are used to compile an outlier time vector $\hat{D}^{(i)}=[\hat{d}_1^{(i)},\hat{d}_2^{(i)},\dots,\hat{d}_{\hat{K}_i}^{(i)}]$ where $\hat{d}_k^{(i)}$ is the observation time, expressed in nanoseconds, measured considering the start time of the current period $T_i$ as the reference time. The outlier time vector can be compiled almost in real-time, with a delay of $v$ packets.
  \item \emph{Watermark identification.} The watermark can be detected in the flow by counting the number of dropping intervals in which we observe at least one outlier. Let $\gamma_i$ be the number of dropping intervals with an outlier identified in the $i$-th time period; if $\gamma_i/K_i$ is greater than a predefined threshold $\beta$, the flow is labelled as watermarked.
\end{itemize}

\section{Invisibility}
\label{sec:invisibility}
A watermark should go unidentified by the adversary, because otherwise she could take some action to prevent the staging server from being detected, say, for example, by interrupting the communication, or in some way preventing the adversary from connecting to the identified staging server so it won't be identified. Invisibility is an important property of a watermark, and it is challenging to obtain. We refer to the definition of statistical invisibility as defined by Iacovazzi~\cite{iacovazzi2017network} saying that ``a watermark is statistically invisible if the difference between the statistical distribution of a watermarked flow and a non-watermarked flow is negligible.''

\subsection{Threat model}
In order to analyze the invisibility of \dw{}, we consider a threat model where an adversary is able to accurately identify the packet loss events in the observed communication. \dw{} pseudo-randomly drops packets according to a modified version of the extended Gilbert model with predefined parameters; although the adversary cannot distinguish between dropped and lost packets, she may suspect the presence of a watermark when the observed statistical behavior differs from a natural behavior. In our model of threat against the watermark's invisibility, the adversary is able to use the watermarker as a black box (thus, she has no knowledge about what is happening internally), and let her traffic traverse it in order to observe the losses and derive the corresponding loss model.
If the packet drop model used by the watermarker exactly fits the loss model of a real bottleneck component, the adversary cannot distinguish between the two and is therefore unable to differentiate watermarked from non-watermarked traffic.
However, when the loss model used for dropping packets only approximates a real behavior, the adversary might suspect the presence of a watermark. This implies that the property of invisibility is strictly related to the goodness of fit of the probabilistic model used to drop packets.

In the next subsection we show that the packet drop statistics of \dw{} overlap with the statistics of a real network component that loses packets because of natural buffer overflow.

\subsection{Evaluation of \dw{}'s degree of invisibility}
We evaluated the goodness of fit and \dw{}'s degree of invisibility based on an empirical study of the loss density and the autocorrelation function (according to the analysis adopted by Yu et al.~\cite{yu2005accuracy}).

Let $\mathcal{W}$ and $\mathcal{M}$ be two statistical models describing the watermarker and the single bottleneck node, respectively. Given the binary vector of packet loss events $\mathbf{B}=[b_1,b_2,\dots,b_{N}]$ observed for traffic going out of the black box, which can be either the watermarker or the bottleneck node, composed of $N$ packets, the adversary aims at recognizing whether $\mathbf{B}$ has been generated by $\mathcal{M}$ or not. This can be cast into a hypothesis test problem, with simple hypothesis $\mathcal{H}=\mathcal{W}$ and $\mathcal{H}=\mathcal{M}$.

The loss density $f_{\mathbf{B}}(k,q)$ is the frequency of $k$ loss events in a block of $q$ events, and the autocorrelation function $\rho_{B}(h)$ for lag $h$ is defined as
\begin{equation}\label{eq:distrib1}
  \rho_{B}(h) = \frac {c_h}{c_0}
\end{equation}
where
\begin{equation}\label{eq:aaa}
  c_h = \frac {1}{N-1}\cdot\sum_{i=1}^{N-i}(b_i-\bar{b})(b_{i+h}-\bar{b}).
\end{equation}

\begin{figure}
\begin{center}
  \includegraphics[width=0.32\textwidth]{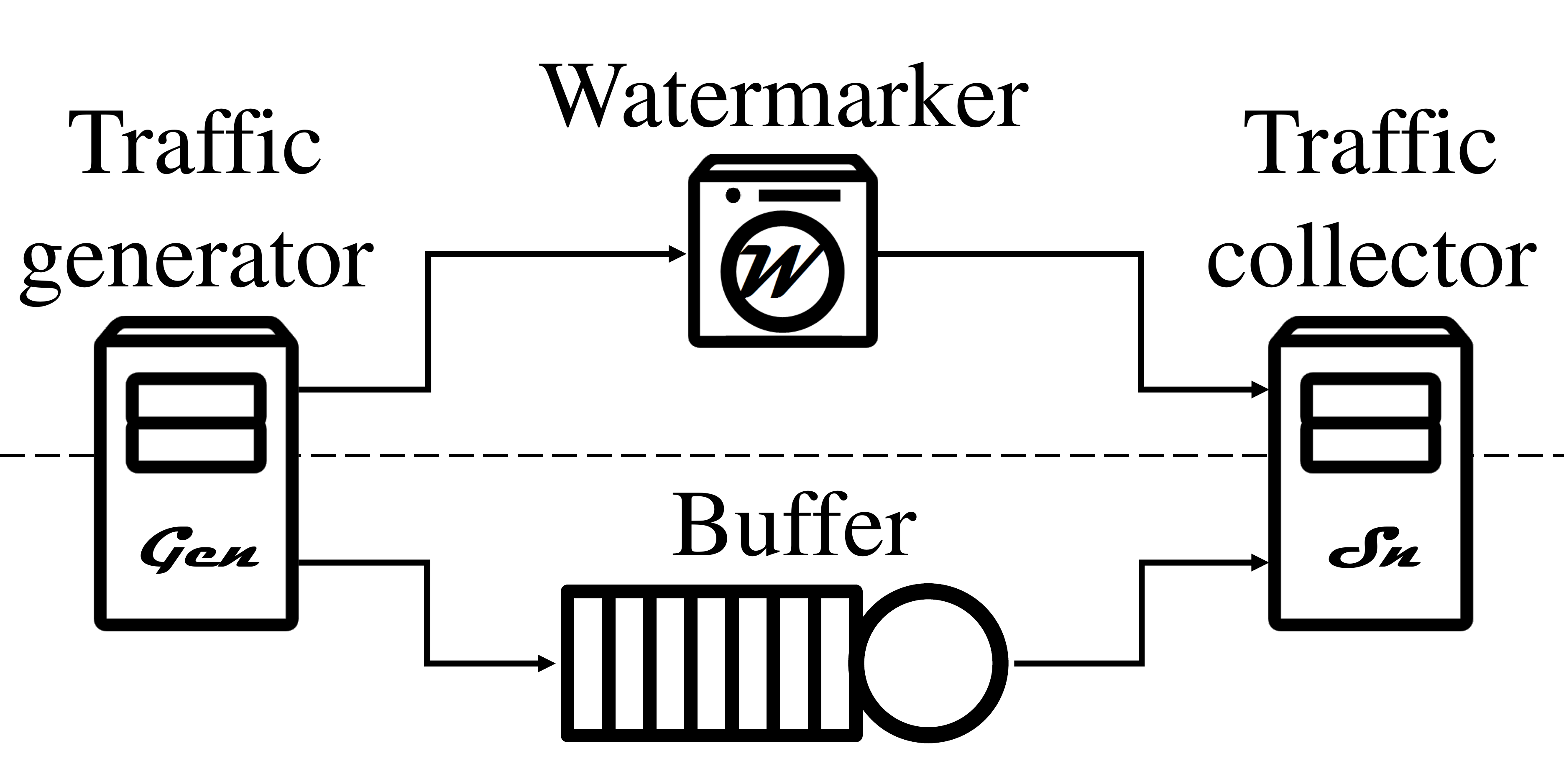}
  \caption{Experimental setup.}\label{fig:bb}
\end{center}
\end{figure}

We measured $f_{\mathbf{B}}(k,q)$ and $\rho_{B}(h)$ for real traffic captured in a controlled experiment. We created a simple network composed of three nodes (Figure~\ref{fig:bb}): 1) a traffic source equipped with IXIA BreakingPoint VE, a commercial traffic generator software capable of generating network traffic at a rate of up to 1~Gbit/s; 2) an intermediate node, acting with the role of either bottleneck or \dw{} watermarker; and 3) a traffic destination node where traffic was collected in order to extract the corresponding binary vector of packet loss events. We configured the source node in order to generate two types of traffic with average transfer rate at $R_{gen} =$ 900~Mbit/s: 1) enterprise traffic composed of a mix of 15 classes (HTTP Video, HTTP Audio, HTTP Text, SIP/RTP Direct Voice Call over TCP, SIP/RTP Direct Voice Call over UDP, SMTP Email, AOL Instant Messenger, DCE RPC, SMB Null Session, SMB Client File Download, NFSv3, PostgreSQL, RTSP, SSH, and FTP), and ``bandwidth'' traffic containing only HTTP and peer-to-peer file sharing, with bandwidth shared evenly (among HTTP, BitTorrent, and eDonkey).

We executed two sets of experiments: one in which the intermediate node was a bottleneck node implemented on a Linux device with a limited egress queue of predetermined size $z=$ 10~pkts, and one in which the intermediate node was the watermarker. 100~GB of traffic was transferred for each experiment. Using the scenario with the bottleneck node, we conducted 11 experiments; the binary vectors extracted from 10 experiments were used to compute the selected metrics, while the last binary vector was used as a training dataset to estimate the probabilities $\{p_{W,k}\}_{0<|k|\leq n}$ to use in the model $\mathcal{W}$. 10 experiments were also conducted using the scenario with the watermarker.

\begin{figure*}
\begin{center}
\subfloat[]{\includegraphics[width=0.32\textwidth]{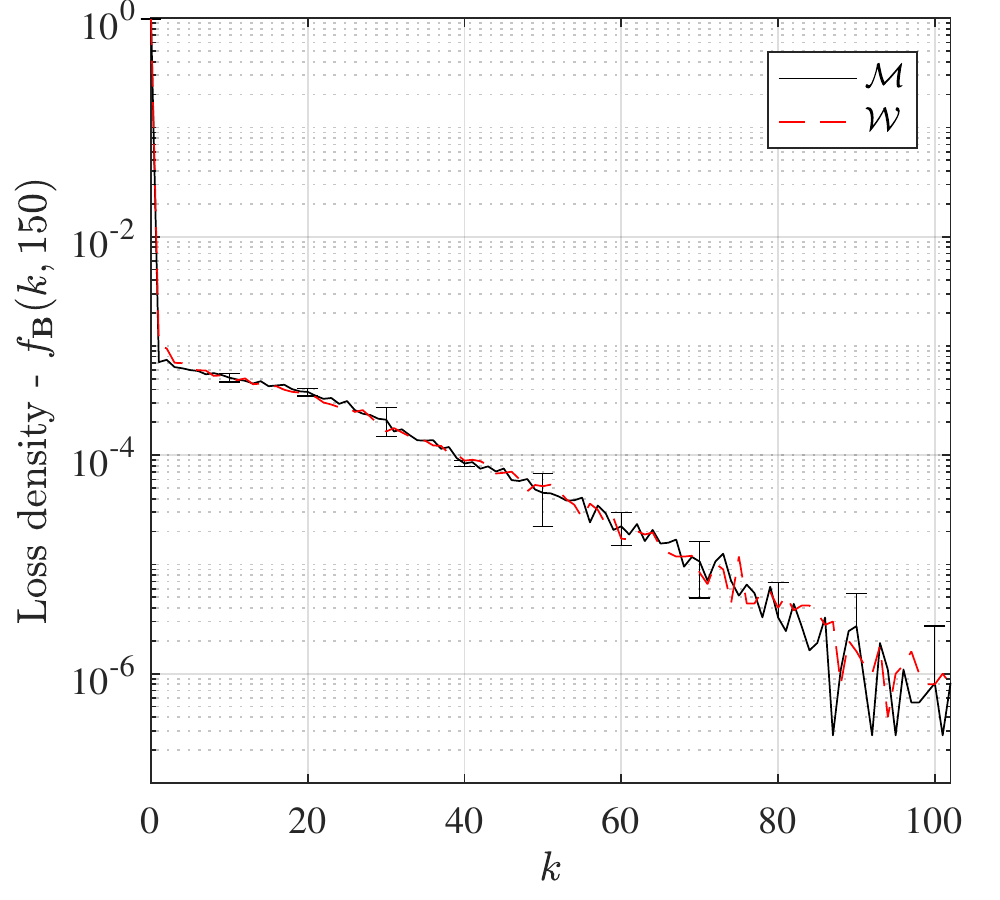}\label{fig:pjn_ent}}
\subfloat[]{\includegraphics[width=0.32\textwidth]{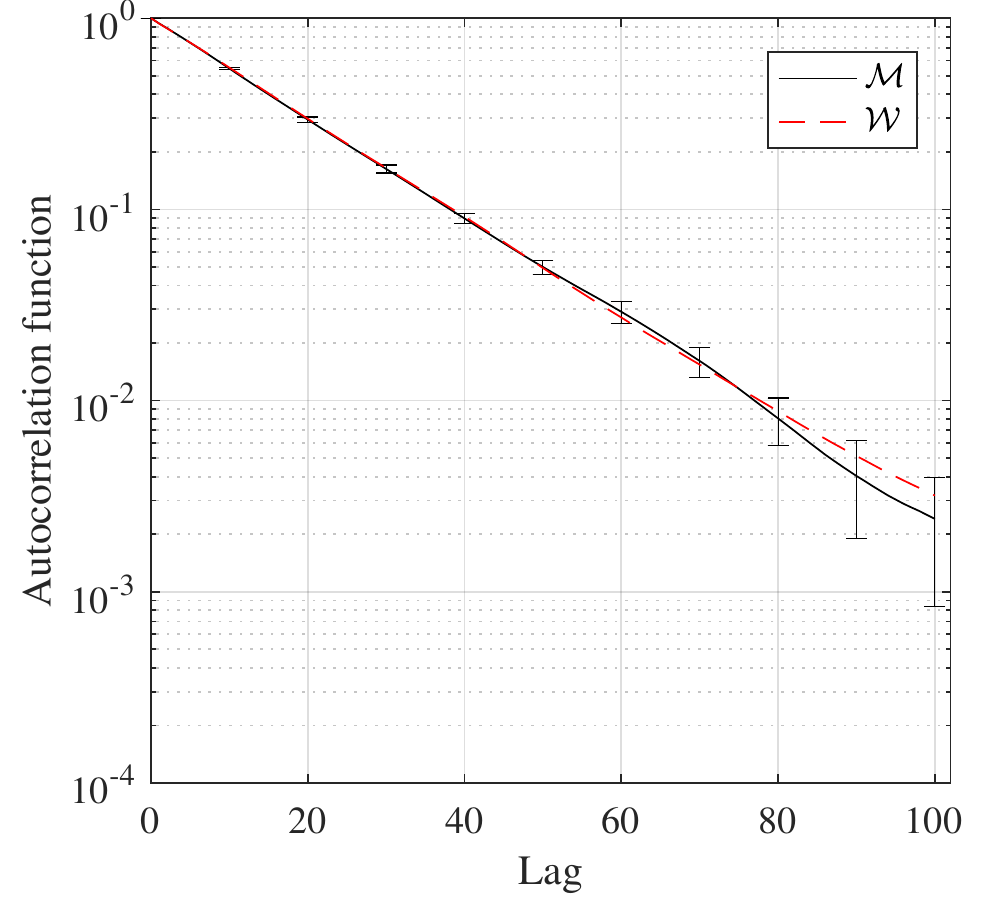}\label{fig:acf_ent}}
\caption{Loss density functions and autocorrelation functions for enterprise traffic.}
\label{fig:enterprise}
\end{center}
\end{figure*}

\begin{figure*}
\begin{center}
\subfloat[]{\includegraphics[width=0.32\textwidth]{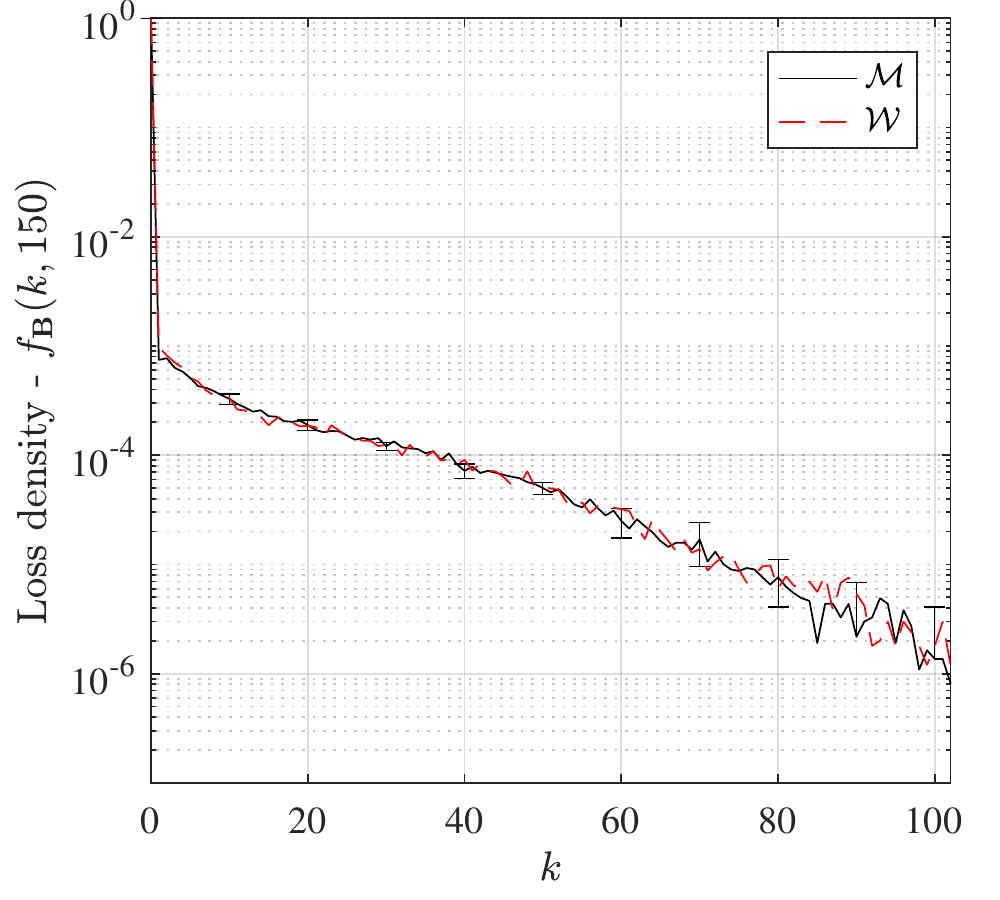}\label{fig:pjn_ban}}
\subfloat[]{\includegraphics[width=0.32\textwidth]{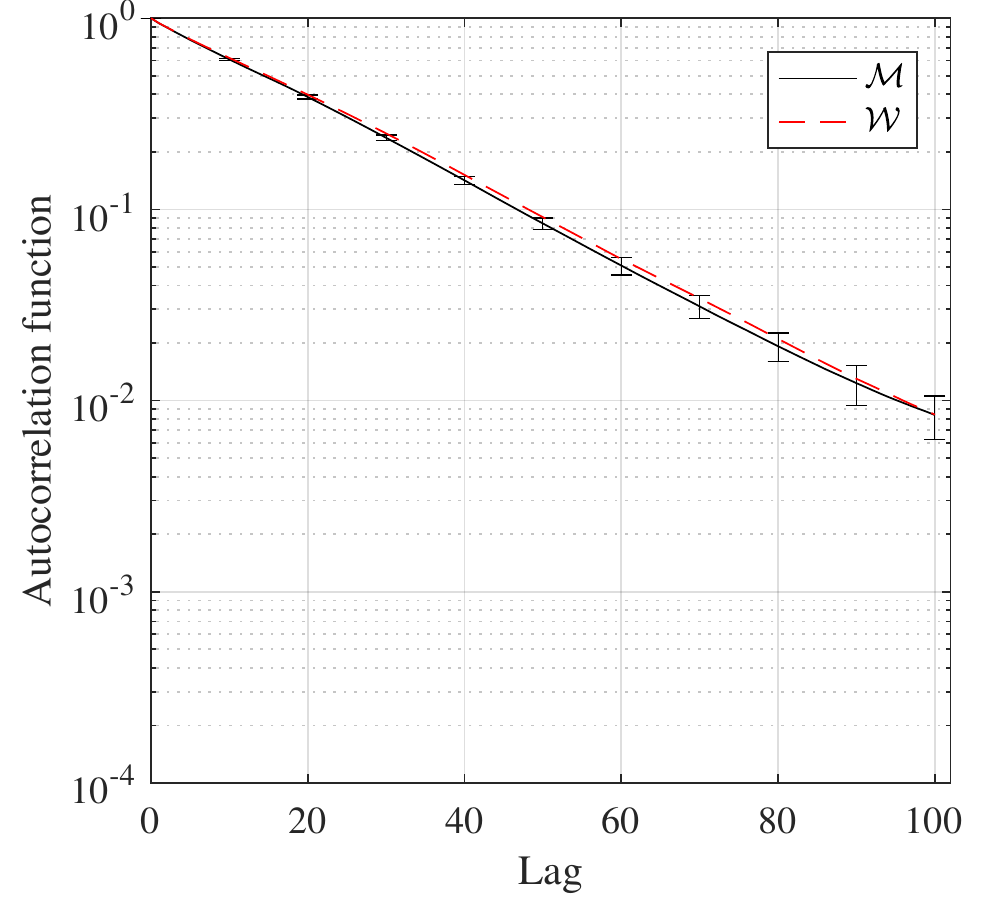}\label{fig:acf_ban}}
\caption{Loss density functions and autocorrelation functions for bandwidth traffic.}
\label{fig:bandwidth}
\end{center}
\end{figure*}

Figures \ref{fig:enterprise} and \ref{fig:bandwidth} provide a comparison of the loss density (for $q=150$) and the autocorrelation function for the two models $\mathcal{W}$ and $\mathcal{M}$ with two compositions of traffic classes. The figures show that the statistics for model $\mathcal{W}$ nearly match those measured for $\mathcal{M}$, and they always stay within the uncertainty level of $\mathcal{M}$.

In order to obtain a numerical measure of the invisibility of \dw{}, we used the Kolmogorov-Smirnov (KS) test to determine whether an observed sample generated by the model $\mathcal{W}$ induces to accept or reject the hypothesis $\mathcal{H}=\mathcal{M}$. The test is based on the cumulative distribution function $F_{B}(k,q)$ defined as
\begin{equation}
F_{\mathbf{B}}(k,q) = \sum_{i=0}^{k}f_{\mathbf{B}}(i,q)
\end{equation}

Let $F_{\mathbf{B}}^{\mathcal{J}}(\cdot)$  be the empirical distribution function of the model $\mathcal{J}$, with $\mathcal{J}\in \{\mathcal{W},\mathcal{M}\}$. In the KS test the hypothesis $\mathcal{H}=\mathcal{M}$ is accepted if
\begin{equation}
\sup_{k}|F_{\mathbf{B}}^{\mathcal{W}}(k,q)-F_{\mathbf{B}}^{\mathcal{M}}(k,q)| < \epsilon
\end{equation}

We conducted the hypothesis test against $\mathbf{B}_\mathcal{M}$ and $\mathbf{B}_\mathcal{W}$ (two sequences of events observed in the two experiments with a bottleneck and a watermarker, respectively). The KS distances obtained for the two types of traffic are listed in Table \ref{tab:ks}. In each case the KS distance is below 0.0009 which corresponds to high confidence (99\%) that the two sequences, $\mathbf{B}_\mathcal{M}$ and $\mathbf{B}_\mathcal{W}$, come from the same distribution. Thus, the watermark injected through \dw{} will be invisible to any third party.

\begin{table}[h]
\centering
\caption{Kolmogronov-Smirnov distances.}\label{tab:ks}
\begin{tabular}{cccc}
    \hline
     Type of traffic & $\bar{b}$ & KS distance  \\
    \hline
    Enterprise &  0.001834 & 0.000873\\
    Bandwidth  &  0.001413 & 0.000733\\
    \hline
\end{tabular}
\end{table}

\section{Performance evaluation}
\label{sec:performance}
We analyze the efficacy of \dw{} based on experiments performed in the wild, with real traffic passing through the Internet. Performance was evaluated for two different network scenarios that are usually used for data exfiltration: 1) a scenario where SSs were implemented as Web proxy servers on Amazon Web Services (AWS), and 2) a scenario where the traffic is forwarded over TOR, the well-known onion routing network.

\subsection{Scenario with Web proxy servers}
\begin{figure*}
\begin{center}
  \includegraphics[width=0.6\textwidth]{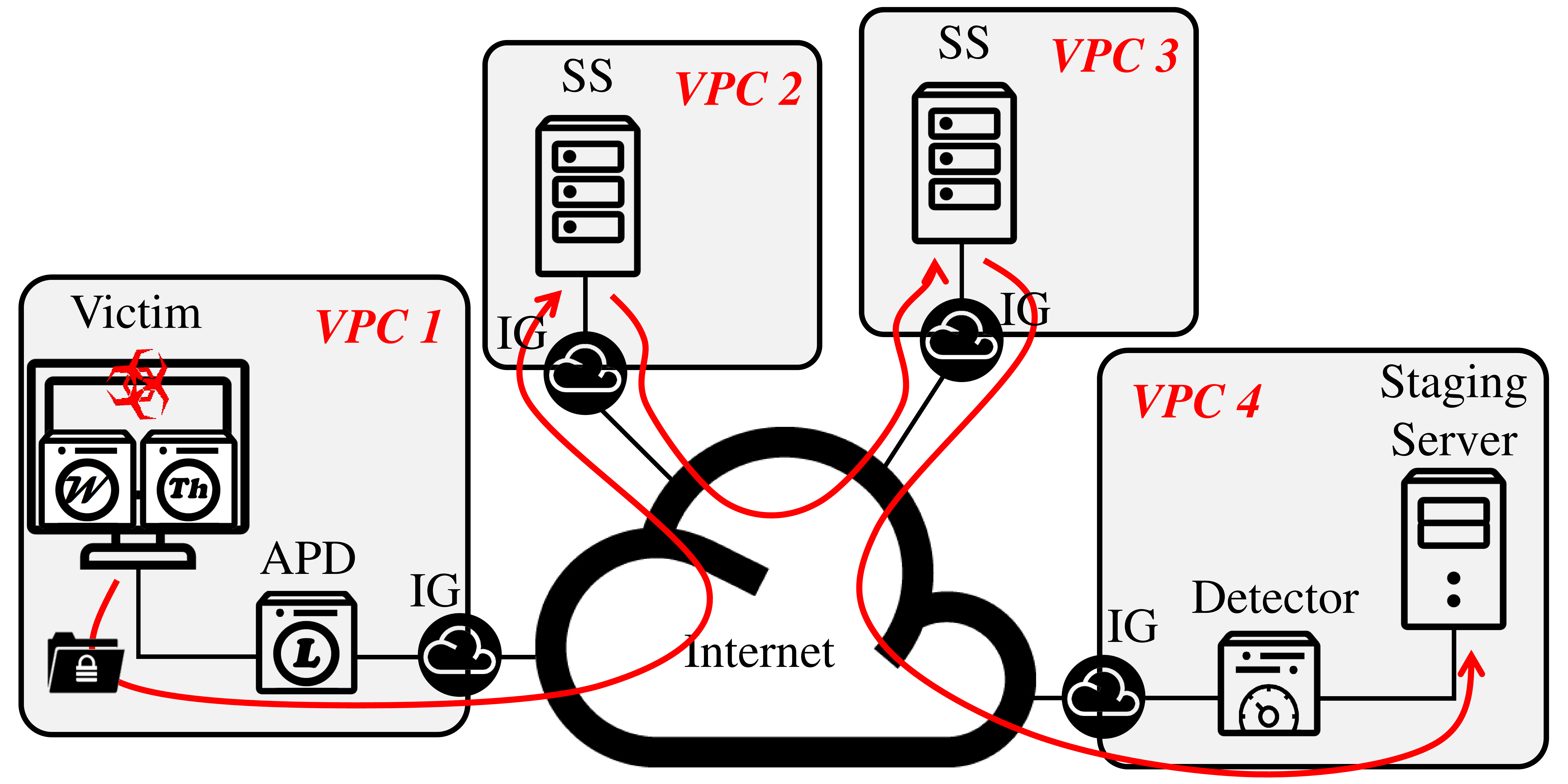}
  \caption{Scenario with Web proxy servers in AWS.}\label{fig:aws-framework}
\end{center}
\end{figure*}

We developed a testing framework in which each node of the network topology is executed in an Amazon Elastic Compute Cloud (Amazon EC2) instance on AWS. Figure~\ref{fig:aws-framework} shows the scheme of the framework. The main components in this architecture are:
\begin{itemize}
  \item \emph{Virtual private cloud (VPC)}: a logically isolated network unit in AWS where one or more EC2 instances can be launched.
  \item \emph{Internet gateway (IG)}: a gateway that interconnects the instances in a VPC with the Internet.
  \item \emph{Victim}: an EC2 instance representing the infected device of a company or a person where sensitive data is stored. For testing purposes the module implementing the watermarker has been installed on this instance. A module which throttles the traffic in order to limit and control the bandwidth used by the malware is also installed on this instance.
  \item \emph{Staging server}: an EC2 instance representing the remote server where the attacker forwards the exfiltrated data.
  \item \emph{Stepping stones (SSs)}: two EC2 instances used in two different VPCs, interposed in the communication from the victim to the staging server.
  \item \emph{Additional packet dropper (APD)}: an EC2 instance that randomly drops packets independently of the watermarker. This is used to test the robustness of \dw{}.
  \item \emph{Detector}: an EC2 instance which sniffs and collects all of the traffic going to the staging server, located in the same VPC as the staging server.
\end{itemize}
The four VPCs were distributed in different geographic regions. A VPC can be launched from one of AWS' 14 regions, distributed around the world; this implies that all of the traffic going from one node to another node passes through the Internet. Our experiment relied upon all of these regions: 10 regions were used to run the SSs, and the remaining four regions were employed to run the VPCs of the victim and the staging server. We used an ``\emph{m4.xlarge}'' instance for the victim and an ``\emph{m4.large}'' for the staging server, both equipped with a Microsoft Windows Server 2012 R2 Base operating system. All other instances were ``\emph{t2.micro}'' equipped with an Ubuntu Server 16.04 LTS.

\subsection{Scenario with onion routing servers}
In this scenario we used the testing framework described in the previous subsection with a few differences: 1) the two SSs running on the EC2 instances were substituted with three onion routers belonging to the real TOR network, 2) the module throttling the traffic installed on the victim's instance and the APD were removed. 

\subsection{Implementation}
\label{sec:implementation}
\subsubsection{Remote administration tool}
Typically, an intruder performs an exfiltration attack by exploiting a remote access Trojan (RAT) which is usually downloaded invisibly on a victim's device within the targeted company network. Once the RAT malware program has been installed, a backdoor is created allowing an attacker to obtain administrative control over the targeted computer. We used a commonly used backdoor malware for Windows systems, generated by Cerberus RAT (a RAT software publicly available on the Internet) and installed on the victim instance; the Cerberus remote controller was installed on the staging server.
\label{sec:framework}

\subsubsection{Stepping stones}
In scenario A, SS implementation is based on SSH protocol. A putty SSH client was used to create two SSH tunnels from the victim to each SS. Proxifier~\cite{proxifier}, a Windows-based proxy software, was used to set up two SOCKS-based proxies in the victim: one to channel Trojan-based TCP connections to the SSH tunnel connecting to the second SS, and the other one to divert the SSH tunnel of the second SS via the SSH tunnel connecting to the first SS. This creates an end-to-end encrypted channel, with one SSH tunnel encapsulated into the other. We set up 20 SSs distributed over 10 different AWS regions. At the beginning of each experiment two SSs were randomly selected by the victim, and the two corresponding SSH tunnels were established.

In scenario B, the application traffic from the victim instance was tunnelled through the onion network using Torifier, a Windows-based torification tool \cite{torifier}. We used the default TOR configuration that uses three relays to build the circuit. At the beginning of each experiment three TOR relays were selected by the victim, and a new circuit was established. In cases in which the selected circuit was exactly the same as the previous experiment, one of the three relays was substituted with a new randomly selected relay.

\subsubsection{Watermarker}
An application conducting the offline and online functions of the watermarker, implemented in C++, was executed on the victim instance. In the online application, Windows Packet Divert (WinDivert)~\cite{windivert}, a packet filter library available for Windows distribution, was used to filter and queue network flow packets from the Windows network stack to the watermarker.
Based on the precomputed dropping sequence, all of the packets observed during dropping time intervals were dropped.

\subsubsection{Network throughput and additional packet loss}
Network throughput can affect the performance of \dw{}. We used the NetLimiter program~\cite{netlimiter} (installed on the victim instance) to throttle the Cerberus traffic and test different values of bandwidth use.

Increasing the packet loss in the network was suggested by Sadeghi et al.~\cite{schulz2014silence} to mitigate covert channels based on packet drops. We tested the robustness of \dw{} against several rates of additional packet loss; NetEm, a Linux facility for traffic control, was used in the Linux instance acting as a APD in order to emulate different network packet loss rates and test the robustness of \dw{}.

\subsubsection{Detector}
For testing purposes, detection was performed offline. Thus, no specific implementation was required in our framework - only an instance to intercept and sniff all of the traffic directed to the staging server was deployed.

\subsection{Numerical results}
\begin{figure*}
\begin{center}
\subfloat[$R =$ 0.5~MB/s]{\includegraphics[width=0.32\textwidth]{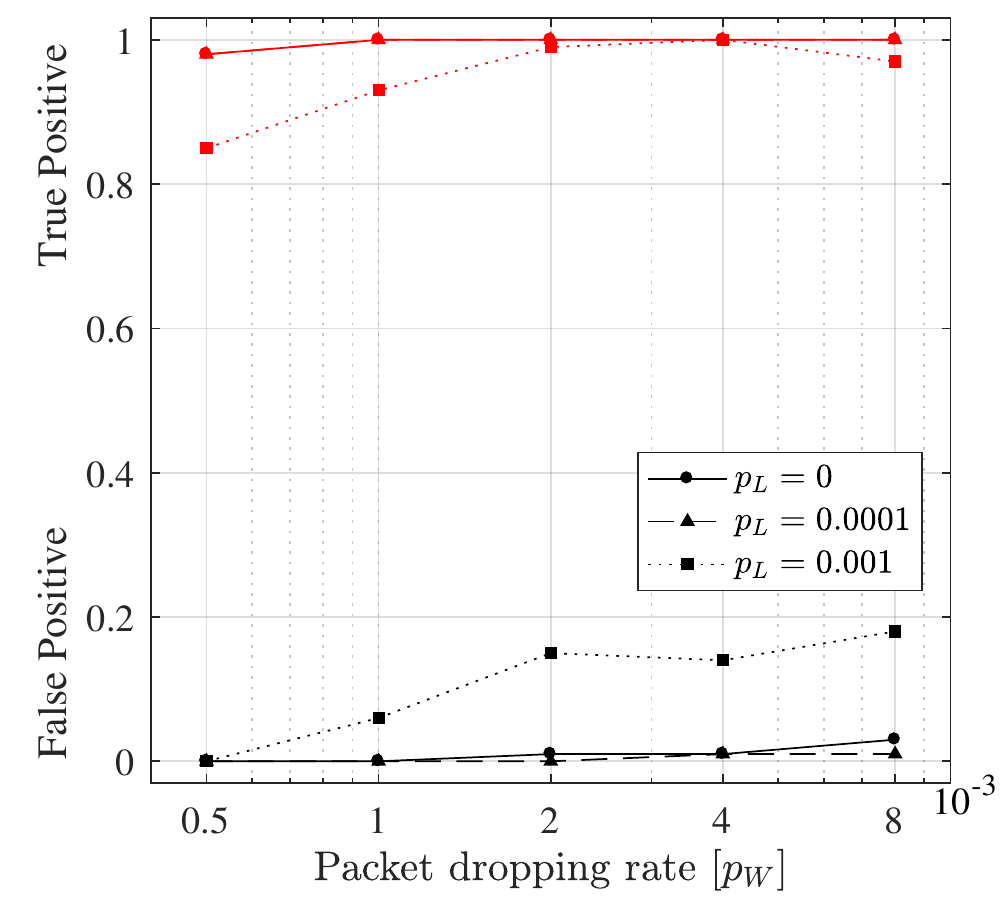}\label{fig:tp05MB4}}
\subfloat[$R =$ 1.0~MB/s]{\includegraphics[width=0.32\textwidth]{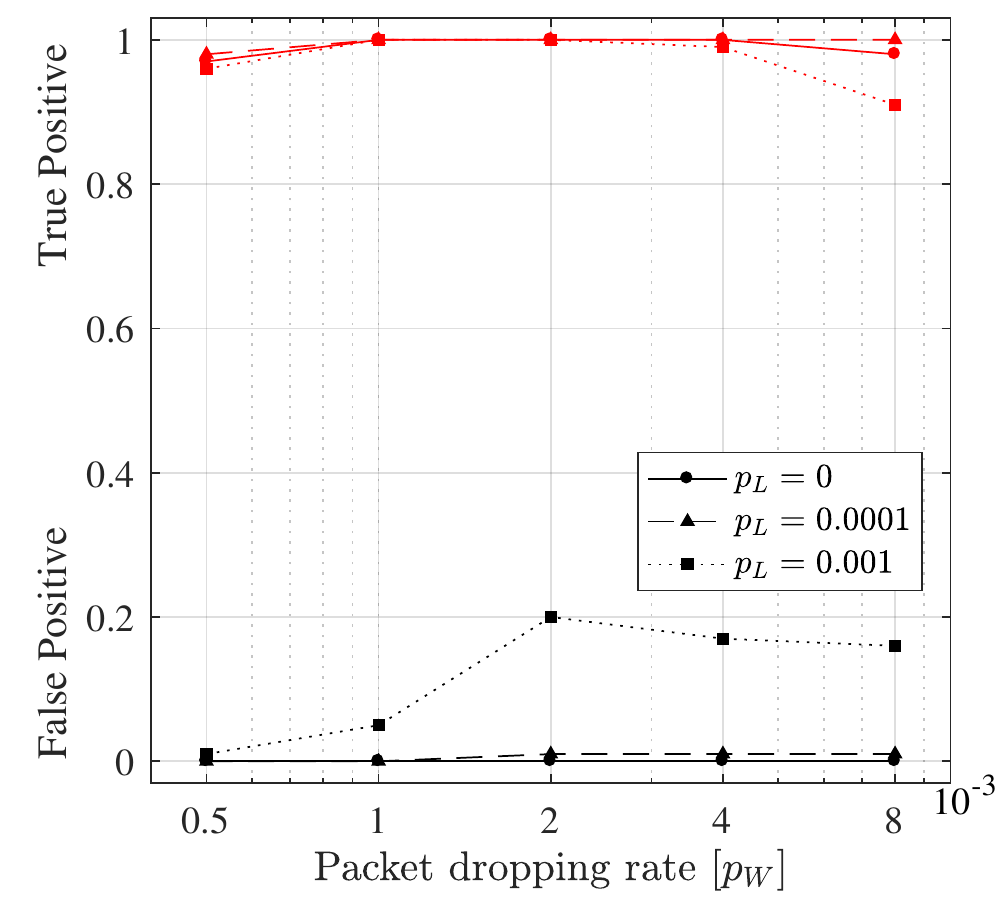}\label{fig:tp1MB4}}
\subfloat[$R =$ 2.2~MB/s]{\includegraphics[width=0.32\textwidth]{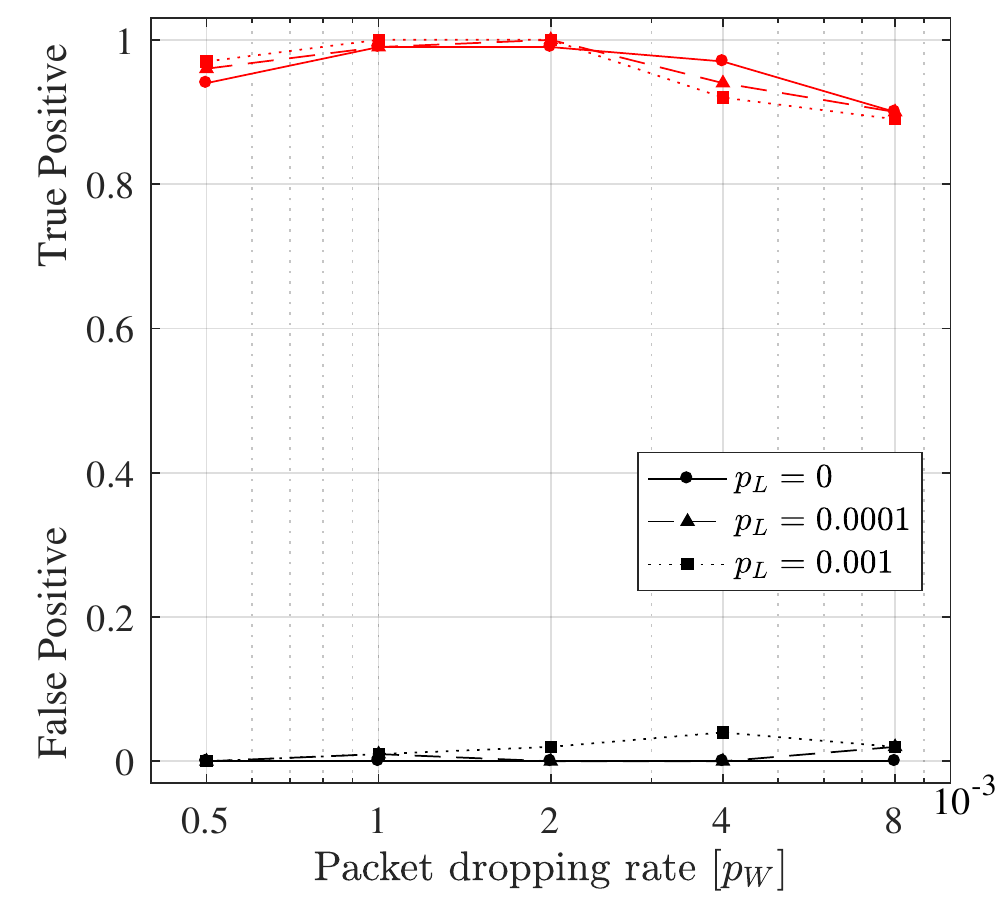}\label{fig:tp2MB4}}
\caption{True and false positive rates in the scenario with Web proxy servers ($\beta = 0.25$).}
\label{fig:TP_FP_4}
\end{center}
\end{figure*}

\begin{figure*}
\begin{center}
\subfloat[$R =$ 0.5~MB/s]{\includegraphics[width=0.32\textwidth]{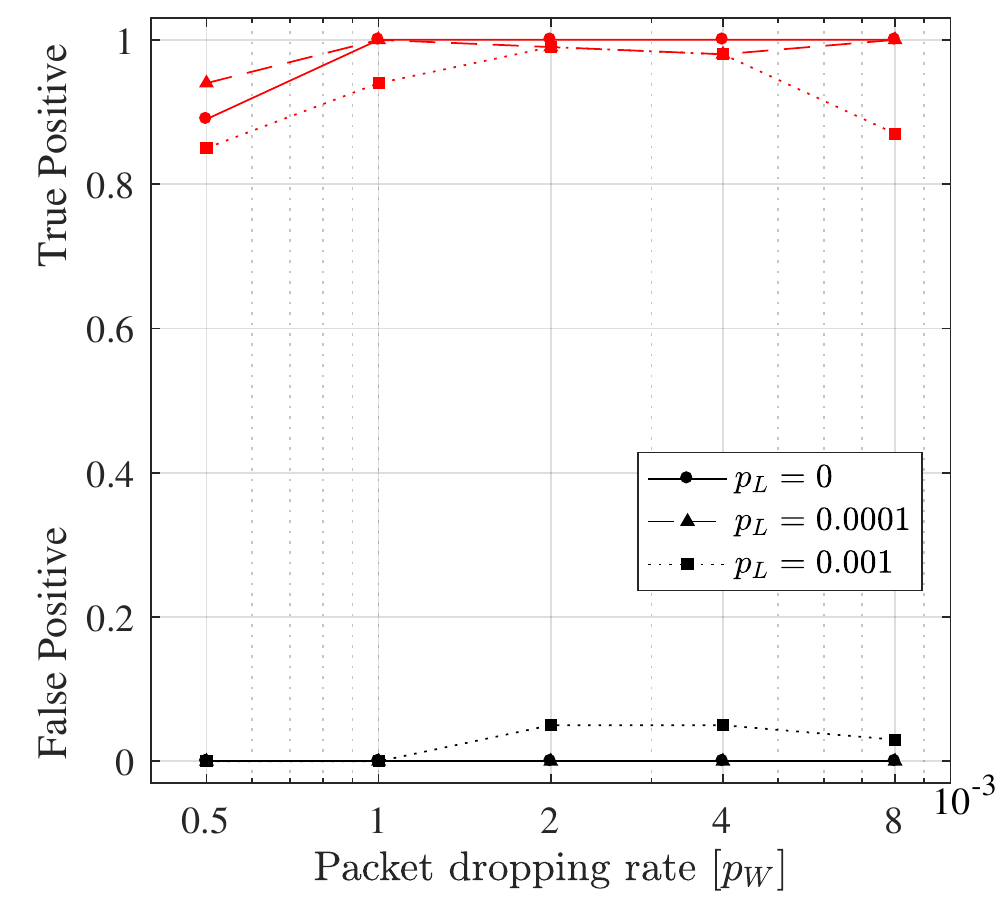}\label{fig:tp05MB6}}
\subfloat[$R =$ 1.0~MB/s]{\includegraphics[width=0.32\textwidth]{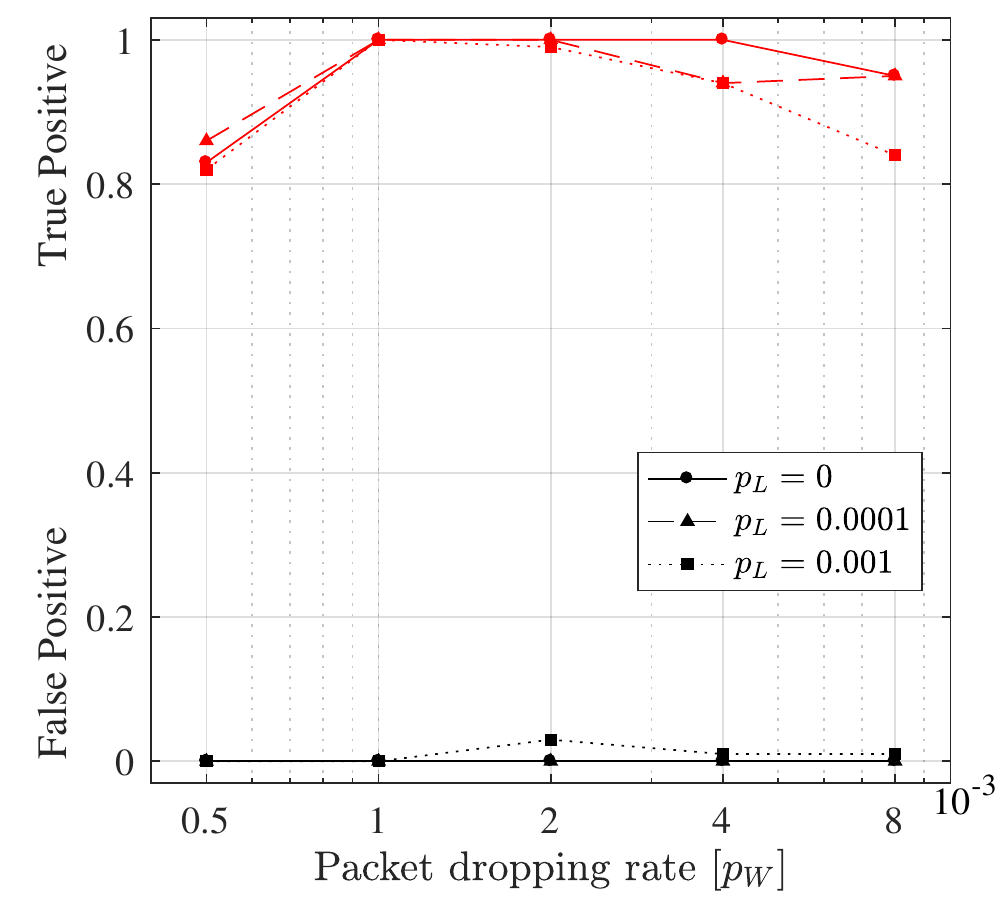}\label{fig:tp1MB6}}
\subfloat[$R =$ 2.2~MB/s]{\includegraphics[width=0.32\textwidth]{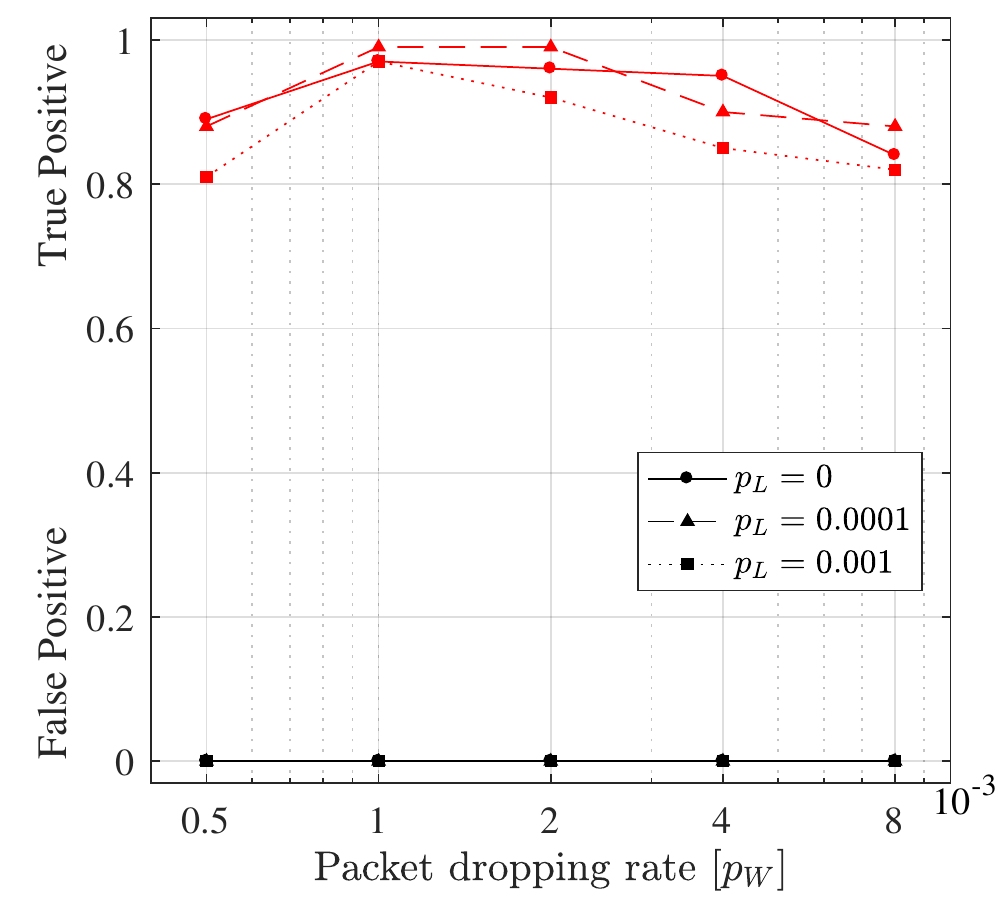}\label{fig:tp2MB6}}
\caption{True and false positive rates in the scenario with Web proxy servers ($\beta = 0.35$).}
\label{fig:TP_FP_6}
\end{center}
\end{figure*}

\dw{}'s accuracy was evaluated by conducting an extensive series of experiments; 7200 experiments on AWS and 500 experiments on TOR were executed, varying values of several parameters: the transfer rate $R$, the packet dropping rate $p_W$, and the additional packet loss rate $p_L$. Each experiment consisted of transferring a file of size 150~MB from the victim to the staging server. We measured the true positive (TP) parameter as the percentage of the watermarked flows correctly classified as watermarked, and the false positive (FP) parameter as the percentage of non-watermarked flows erroneously classified as watermarked.

A training phase was performed on a training dataset composed of 50 traces in order to test several values of the outlier threshold $\alpha$ and comparison window $v$, and to select the values to use in the evaluation of the system. After the training phase, we selected $\alpha=0.8$ and $v=$ 300~pkts.
The selection of the $\{p_{W,k}\}_{0<|k|\leq n}$ to use in the model $\mathcal{W}$ was made based on a training trace made of 100~GB of traffic, captured in the bottleneck setup described in Section \ref{sec:invisibility}.

Figures~\ref{fig:TP_FP_4} and~\ref{fig:TP_FP_6} show the TP and FP rates obtained in our experiments in the scenario with Web proxy servers, for three values of transfer rate ($R=$0.5, 1.0, and 2.2~MB/s) and two values of $\beta$ (0.25 and 0.35). On each graph, TP (red) and FP (black) rates are plotted for different values of $p_L$ and with various packet dropping rates $p_W$. Each point on the curves corresponds to an average value computed over 100 experiments. FP rates were evaluated by testing the detector with both non-watermarked traces and traces watermarked with a wrong seed.
\begin{figure*}
\begin{center}
\subfloat[$R =$ 0.5~MB/s]{\includegraphics[width=0.32\textwidth]{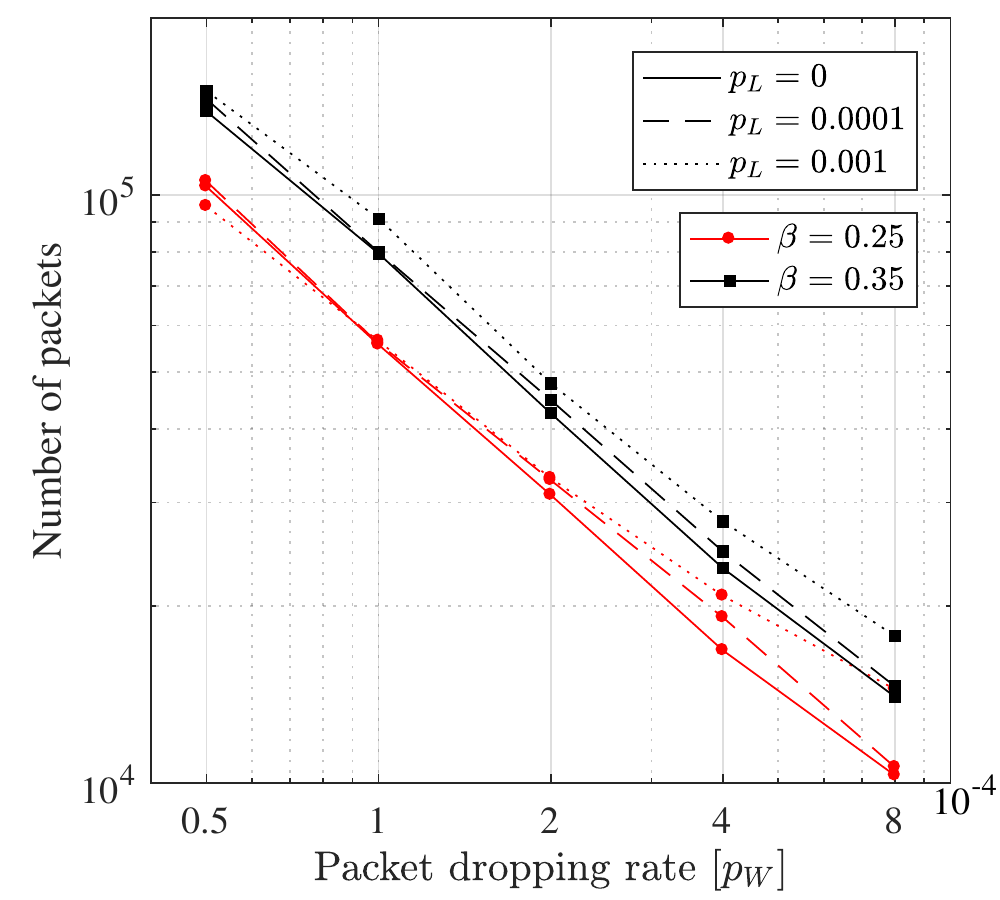}\label{fig:np05MB}}
\subfloat[$R =$ 1.0~MB/s]{\includegraphics[width=0.32\textwidth]{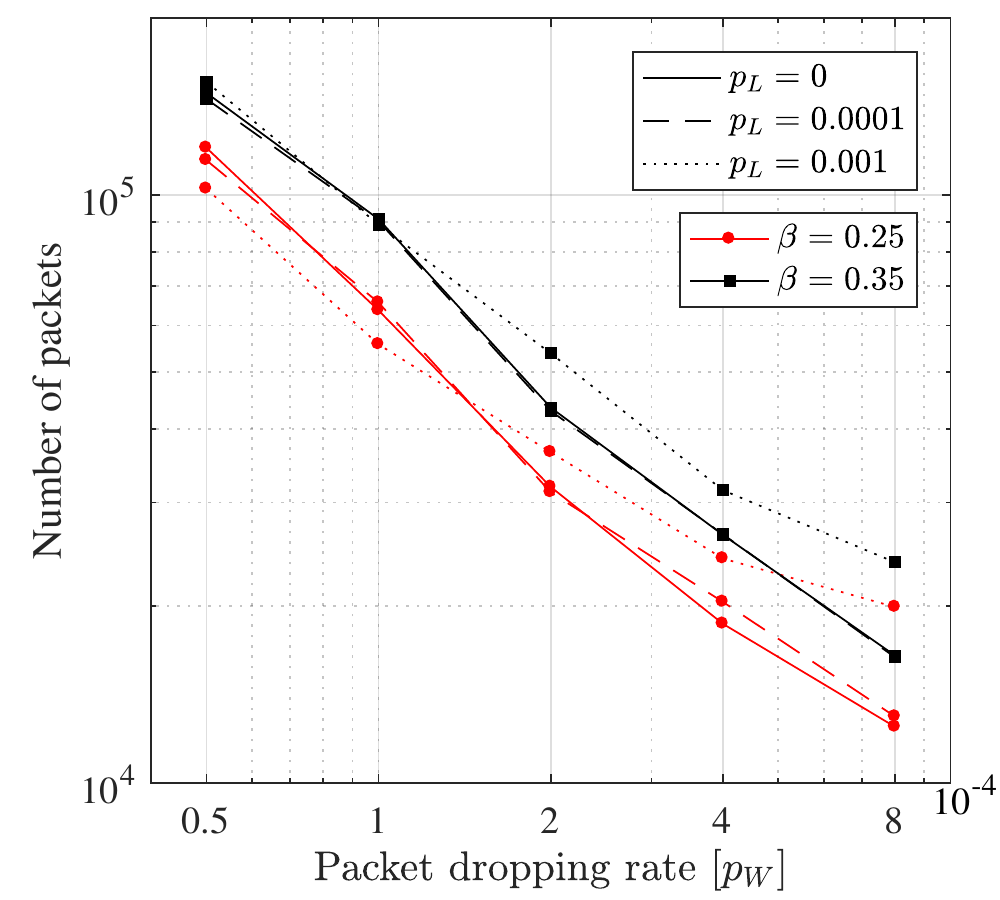}\label{fig:np1MB}}
\subfloat[$R =$ 2.2~MB/s]{\includegraphics[width=0.32\textwidth]{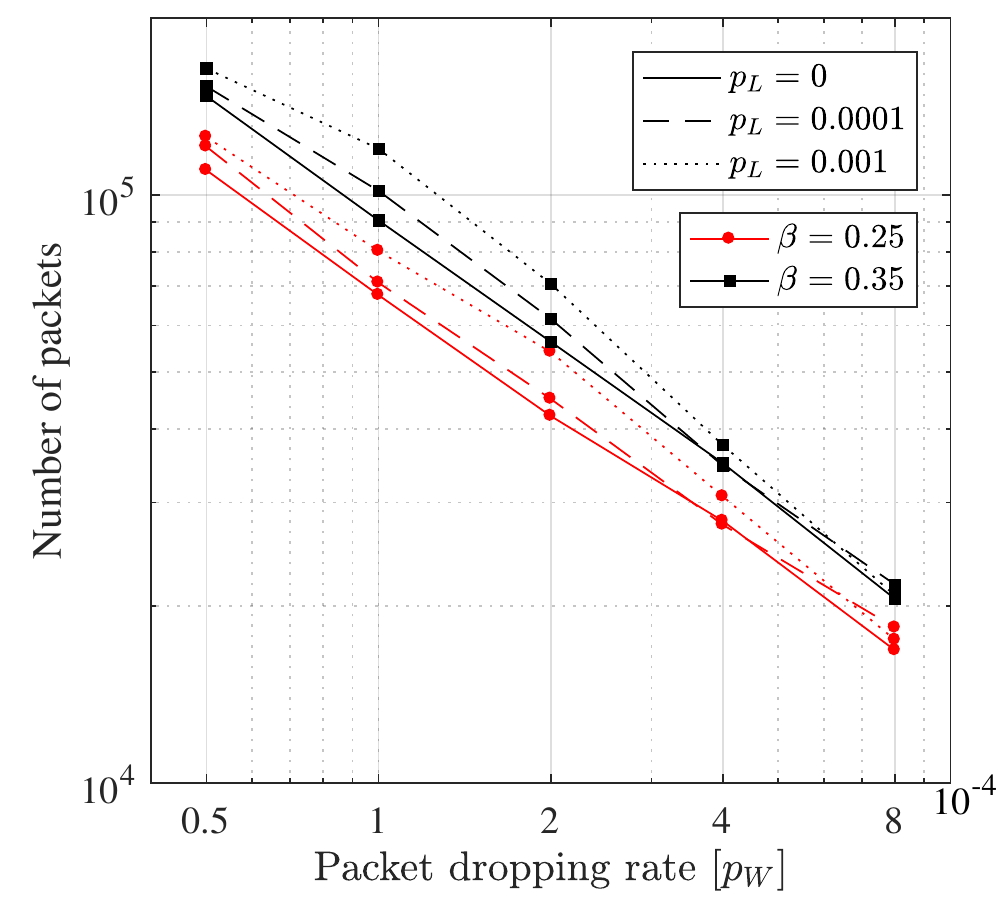}\label{fig:np2MB}}
\caption{Number of packets required to detect a watermark ($\beta = 0.25$ in red and $\beta = 0.35$ in black).}
\label{fig:number_4}
\end{center}
\end{figure*}

As can be seen in the graphs, the detection algorithm is able to correctly detect watermarks, achieving very high TP rates (over 95\% in most cases) and low FP rates (below 5\%). Although variations on transfer rates did not significantly affect performance, 
we observed a minor deterioration in TP for cases in which transfer rate is 2.2~MB/s, nevertheless it is still effective to detect watermarks with few errors. When packet loss is greater than a specific threshold, a significant amount of noise is added to the sequence of IPDs which very slightly hinders the outlier detection function. For the same reason, TP rates also worsened as the combination of packet loss and packet drop frequencies increased. In addition, we observed a slight deterioration in the TP rate for $\beta = 0.35$. This is due to the fact that increasing the level of the detection threshold $\beta$ reduces the implicit redundancy inside the watermark, which affects \dw{}'s TP, but at the same time it drastically reduces FP rates. Thus, the detection system is highly effective even at a higher threshold with less error.
A slight downturn in TP rates can also be observed for $p_W=0.5\cdot 10^{-3}$; this is explained by the fact that in this scenario not enough packets are dropped before the file transfer ends.
FP rates increase in a scenario with high packet loss rate and low throughput when we consider a threshold $\beta = 0.25$; nevertheless FP rates are below 5\% in all of the other cases.
Therefore, we can state that, for practical implementation, watermarks can be detected with almost 100\% TP and 0\% FP rates with the fine-tuning of the system parameters according to the knowledge of network loss behavior, even with the presence of a mitigation technique.

Figures~\ref{fig:number_4} 
show the number of packets required to detect the watermark for three values of transfer rate ($R=$ 0.5, 1.0, and 2.2~MB/s). On each graph, curves are plotted for three values of $p_L$, and for $\beta=0.25$ (red) and $\beta=0.35$ (black), by varying the packet dropping probability $p_W$. Each point on the curves corresponds to an average value computed over all of the experiments that resulted in the correct identification of a watermark. The number of packets needed to identify the watermark ranges from $10^4$ to $1.5\cdot 10^5$. It is no surprise that in all of the cases the number of packets required for detection decreases linearly as the packet dropping rate increases.

\begin{figure}
\begin{center}
  \includegraphics[width=0.32\textwidth]{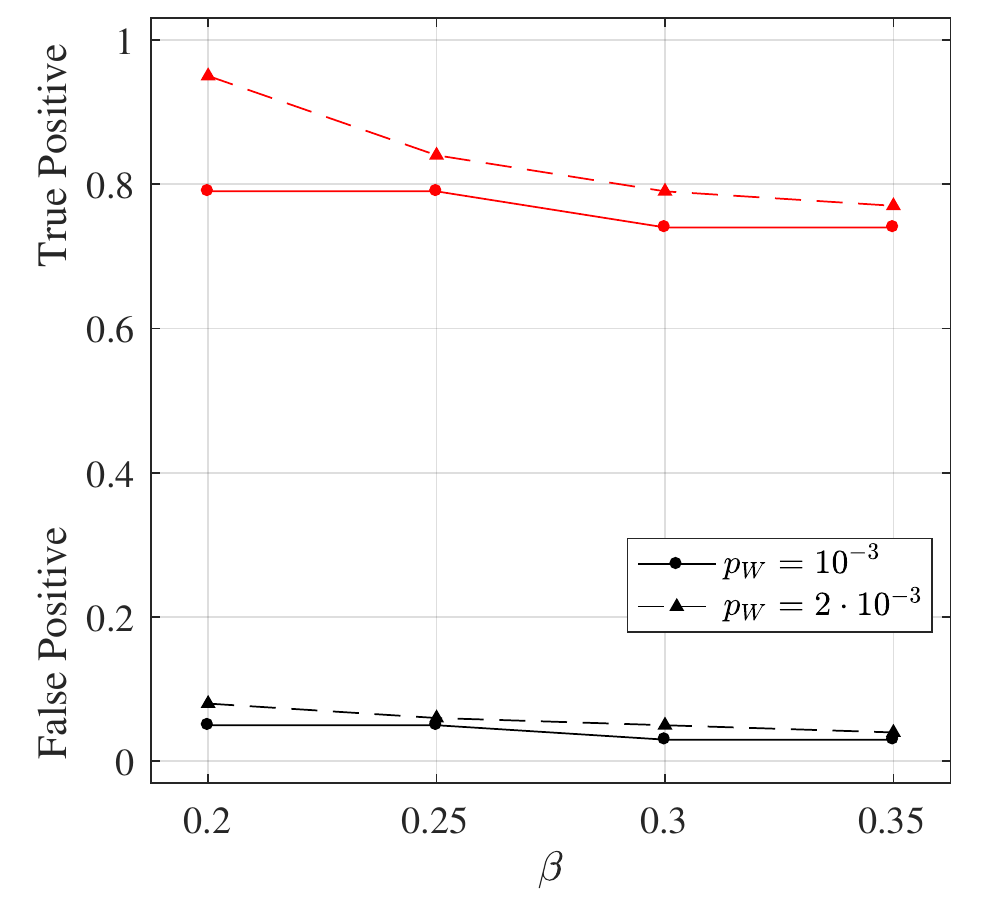}
  \caption{True and false positive rates in the scenario with onion routing servers.}\label{fig:tor1}
\end{center}
\end{figure}

We also tested \dw{} in a scenario with the TOR network. Even though TOR is not optimal for performing the transfer of massive amount of data, testing the watermarking system in a scenario with onion routing servers allows us to stress robustness in the presence of a significant amount of noise that is largely due to relay instability, large end-to-end delay, and large jitter. Figure~\ref{fig:tor1}, shows the TP and FP rates for two values of $p_W$ by varying the threshold $\beta$. Despite the slight decrease in performance, we can detect the watermark in 95\% of cases (best instance), with an FP rate below 10\%.

\section{Discussion and challenges}
\label{sec:discussion}
Unlike the other watermarking algorithms known in the literature, \dw{} has the following properties: 1) it is invisible to the adversary; 2) it is effective, even with a high transfer rate;
and 3) it is effective against traffic passing through the TOR network.

Dropping packets on a pseudo-random basis implies that deterministic analysis would not provide any evidence of the watermark. Additionally, statistical analysis would be incapable of this as well, because the packet loss behavior induced by \dw{} reflects a natural behavior of loss in the network.

The extensive number of experiments performed showed that \dw{} is effective in a variety of scenarios, with different network conditions, even in the presence of different TCP stack implementations. The robustness of \dw{} was also demonstrated in a scenario involving an attacker that intentionally drops packets with the aim of obfuscating the watermark.

One of the \dw{}'s characteristics is the ability to take advantage of the network features, the operation of the SS, and the interaction with network protocols. Nevertheless, as previously stated, the way the SS handles the traffic may significantly affect the incisiveness and detectability of any watermark. For instance, a timing-based algorithm needs the SS to work seamlessly, in order to safeguard the temporal patterns. If an SS uses a ``store and forward'' method, in which received data is buffered for a period of time before being forwarded to the next hop, then all of the watermarking algorithms (including \dw{}) cannot work correctly.

Another limitation is that \dw{} is ineffective for short-lived or interactive flows, because the $p_W$ must be low enough to ensure that 1) packet dropping does not affect the throughput, and 2) TP rates are sufficiently high.

\section{Conclusion}
\label{sec:conclusion}
In this paper we proposed a new watermarking algorithm for tracing data exfiltration attacks. \dw{} is an algorithm that has two main characteristics that differentiate it from other existing solutions for the network traceback problem. First, \dw{}'s embedding algorithm is based on a new paradigm to impress a watermark within a network flow that takes advantage of a network's reaction to packet loss. We have shown that dropping a few selected packets of a flow allows a timing-based watermark to be embedded into the flow.
Second, the watermark embedded by \dw{} is completely invisible to the adversary. The invisibility is due to the fact that the time alteration generated by an artificially dropped packet is the same as that of a packet that is naturally lost. In addition, because the statistical behavior of the loss pattern induced by \dw{} fits the loss behavior of a real bottleneck node, an adversary cannot distinguish between a watermark embedded by \dw{} and a natural loss pattern in  the network. Our experimental results showed that \dw{} achieves very high TP rates and very low FP rates even in realistic scenarios where traffic passes through Web proxy servers on AWS or an anonymous network like TOR.

%\bibliographystyle{plain}
%\bibliography{WMsurveybib}

\begin{thebibliography}{10}

\bibitem{netlimiter}
Netlimiter.
\newblock \url{https://www.netlimiter.com/}.
\newblock Accessed: 2016-09-01.

\bibitem{torifier}
Torifier.
\newblock \url{http://www.torifier.com/}.
\newblock Accessed: 2016-09-01.

\bibitem{windivert}
Windows packet divert (windivert).
\newblock \url{https://reqrypt.org/windivert.html}.
\newblock Accessed: 2016-09-01.

\bibitem{proxifier}
Windows proxifier.
\newblock \url{https://www.proxifier.com/}.
\newblock Accessed: 2016-09-01.

\bibitem{arp2015torben}
Daniel Arp, Fabian Yamaguchi, and Konrad Rieck.
\newblock Torben: A practical side-channel attack for deanonymizing tor
  communication.
\newblock In {\em Proceedings of the 10th ACM Symposium on Information,
  Computer and Communications Security}, pages 597--602. ACM, 2015.

\bibitem{bertino2011towards}
Elisa Bertino and Gabriel Ghinita.
\newblock Towards mechanisms for detection and prevention of data exfiltration
  by insiders: keynote talk paper.
\newblock In {\em Proceedings of the 6th ACM Symposium on Information, Computer
  and Communications Security}, pages 10--19. ACM, 2011.

\bibitem{binde2011assessing}
Beth Binde, Russ McRee, and Terrence~J O’Connor.
\newblock Assessing outbound traffic to uncover advanced persistent threat.
\newblock {\em SANS Institute. Whitepaper}, 2011.

\bibitem{buchholz2002providing}
Florian~P Buchholz and Clay Shields.
\newblock Providing process origin information to aid in network traceback.
\newblock In {\em USENIX Annual Technical Conference, General Track}, pages
  261--274, 2002.

\bibitem{chan2013revisiting}
Eric Chan-Tin, Jiyoung Shin, and Jiangmin Yu.
\newblock Revisiting circuit clogging attacks on tor.
\newblock In {\em Availability, Reliability and Security (ARES), 2013 Eighth
  International Conference on}, pages 131--140. IEEE, 2013.

\bibitem{choi2003network}
Yang-Seo Choi, Dong-il Seo, Sung-Won Sohn, and Sang-Ho Lee.
\newblock Network-based real-time connection traceback system (nrcts) with
  packet marking technology.
\newblock {\em Computational Science and Its Applications ICCSA 2003}, pages
  31--40, 2003.

\bibitem{deweese2009capability}
Steve DeWeese.
\newblock {\em Capability of the People's Republic of China (PRC) to conduct
  cyber warfare and computer network exploitation}.
\newblock DIANE Publishing, 2009.

\bibitem{edman2009anonymity}
Matthew Edman and B{\"u}lent Yener.
\newblock On anonymity in an electronic society: A survey of anonymous
  communication systems.
\newblock {\em ACM Computing Surveys (CSUR)}, 42(1):5, 2009.

\bibitem{giani2006data}
Annarita Giani, Vincent~H Berk, and George~V Cybenko.
\newblock Data exfiltration and covert channels.
\newblock In {\em Defense and Security Symposium}, pages 620103--620103.
  International Society for Optics and Photonics, 2006.

\bibitem{giura2012context}
Paul Giura and Wei Wang.
\newblock A context-based detection framework for advanced persistent threats.
\newblock In {\em Cyber Security (CyberSecurity), 2012 International Conference
  on}, pages 69--74. IEEE, 2012.

\bibitem{gong2012invisible}
Xun Gong, Mavis Rodrigues, and Negar Kiyavash.
\newblock Invisible flow watermarks for channels with dependent substitution
  and deletion errors.
\newblock In {\em Acoustics, Speech and Signal Processing (ICASSP), 2012 IEEE
  International Conference on}, pages 1773--1776. IEEE, 2012.

\bibitem{gong2013invisible}
Xun Gong, Mavis Rodrigues, and Negar Kiyavash.
\newblock Invisible flow watermarks for channels with dependent substitution,
  deletion, and bursty insertion errors.
\newblock {\em Information Forensics and Security, IEEE Transactions on},
  8(11):1850--1859, 2013.

\bibitem{hamadeh2006taxonomy}
Ihab Hamadeh and George Kesidis.
\newblock A taxonomy of internet traceback.
\newblock {\em International Journal of Security and Networks}, 1(1-2):54--61,
  2006.

\bibitem{houmansadr2011swirl}
Amir Houmansadr and Nikita Borisov.
\newblock Swirl: A scalable watermark to detect correlated network flows.
\newblock In {\em NDSS}, 2011.

\bibitem{houmansadr2013botmosaic}
Amir Houmansadr and Nikita Borisov.
\newblock Botmosaic: Collaborative network watermark for the detection of
  irc-based botnets.
\newblock {\em Journal of Systems and Software}, 86(3):707--715, 2013.

\bibitem{houmansadr2009rainbow}
Amir Houmansadr, Negar Kiyavash, and Nikita Borisov.
\newblock Rainbow: A robust and invisible non-blind watermark for network
  flows.
\newblock In {\em NDSS}, 2009.

\bibitem{iacovazzi2017network}
Alfonso Iacovazzi and Yuval Elovici.
\newblock Network flow watermarking: A survey.
\newblock {\em IEEE Communications Surveys \& Tutorials}, 19(1):512--530, 2017.

\bibitem{jia2013blind}
Weijia Jia, Fung~Po Tso, Zhen Ling, Xinwen Fu, Dong Xuan, and Wei Yu.
\newblock Blind detection of spread spectrum flow watermarks.
\newblock {\em Security and Communication Networks}, 6(3):257--274, 2013.

\bibitem{kiyavash2008multi}
Negar Kiyavash, Amir Houmansadr, and Nikita Borisov.
\newblock Multi-flow attacks against network flow watermarking schemes.
\newblock In {\em USENIX Security Symposium}, pages 307--320, 2008.

\bibitem{lee2013clustering}
Martin Lee and Darren Lewis.
\newblock Clustering disparate attacks: mapping the activities of the advanced
  persistent threat.
\newblock {\em Last accessed June}, 26, 2013.

\bibitem{li2011evidence}
Frankie Li, Anthony Lai, and Ddl Ddl.
\newblock Evidence of advanced persistent threat: A case study of malware for
  political espionage.
\newblock In {\em Malicious and Unwanted Software (MALWARE), 2011 6th
  International Conference on}, pages 102--109. IEEE, 2011.

\bibitem{li2004large}
Jun Li, Minho Sung, Jun Xu, and Li~Li.
\newblock Large-scale ip traceback in high-speed internet: Practical techniques
  and theoretical foundation.
\newblock In {\em Security and Privacy, 2004. Proceedings. 2004 IEEE Symposium
  on}, pages 115--129. IEEE, 2004.

\bibitem{lin2012new}
Zi~Lin and Nicholas Hopper.
\newblock New attacks on timing-based network flow watermarks.
\newblock In {\em USENIX Security Symposium}, pages 381--396, 2012.

\bibitem{ling2013novel}
Zhen Ling, Xinwen Fu, Weijia Jia, Wei Yu, Dong Xuan, and Junzhou Luo.
\newblock Novel packet size-based covert channel attacks against anonymizer.
\newblock {\em Computers, IEEE Transactions on}, 62(12):2411--2426, 2013.

\bibitem{liu2009sidd}
Yali Liu, Cherita Corbett, Ken Chiang, Rennie Archibald, Biswanath Mukherjee,
  and Dipak Ghosal.
\newblock Sidd: A framework for detecting sensitive data exfiltration by an
  insider attack.
\newblock In {\em System Sciences, 2009. HICSS'09. 42nd Hawaii International
  Conference on}, pages 1--10. IEEE, 2009.

\bibitem{luo2012interval}
Junzhou Luo, Xiaogang Wang, and Ming Yang.
\newblock An interval centroid based spread spectrum watermarking scheme for
  multi-flow traceback.
\newblock {\em Journal of Network and Computer Applications}, 35(1):60--71,
  2012.

\bibitem{luo2010secrecy}
Xiapu Luo, Junjie Zhang, Roberto Perdisci, and Wenke Lee.
\newblock On the secrecy of spread-spectrum flow watermarks.
\newblock {\em Computer Security--ESORICS 2010}, pages 232--248, 2010.

\bibitem{luo2011exposing}
Xiapu Luo, Peng Zhou, Junjie Zhang, Roberto Perdisci, Wenke Lee, and Rocky~KC
  Chang.
\newblock Exposing invisible timing-based traffic watermarks with backlit.
\newblock In {\em Proceedings of the 27th Annual Computer Security Applications
  Conference}, pages 197--206. ACM, 2011.

\bibitem{mazurczyk2016information}
Wojciech Mazurczyk, Steffen Wendzel, Sebastian Zander, Amir Houmansadr, and
  Krzysztof Szczypiorski.
\newblock {\em Information hiding in communication networks: Fundamentals,
  mechanisms, applications, and countermeasures}.
\newblock John Wiley \& Sons, 2016.

\bibitem{mitropoulos2005network}
Sarandis Mitropoulos, Dimitrios Patsos, and Christos Douligeris.
\newblock Network forensics: towards a classification of traceback mechanisms.
\newblock In {\em Security and Privacy for Emerging Areas in Communication
  Networks, 2005. Workshop of the 1st International Conference on}, pages
  9--16. IEEE, 2005.

\bibitem{peng2005active}
Pai Peng, Peng Ning, Douglas~S Reeves, and Xinyuan Wang.
\newblock Active timing-based correlation of perturbed traffic flows with chaff
  packets.
\newblock In {\em Distributed Computing Systems Workshops, 2005. 25th IEEE
  International Conference on}, pages 107--113. IEEE, 2005.

\bibitem{pyun2007tracing}
Young~June Pyun, Young~Hee Park, Xinyuan Wang, Douglas~S Reeves, and Peng Ning.
\newblock Tracing traffic through intermediate hosts that repacketize flows.
\newblock In {\em INFOCOM 2007. 26th IEEE International Conference on Computer
  Communications. IEEE}, pages 634--642. IEEE, 2007.

\bibitem{ramsbrock2008first}
Daniel Ramsbrock, Xinyuan Wang, and Xuxian Jiang.
\newblock A first step towards live botmaster traceback.
\newblock In {\em Recent Advances in Intrusion Detection}, pages 59--77.
  Springer Berlin Heidelberg, 2008.

\bibitem{sanneck1999framework}
Henning~A Sanneck and Georg Carle.
\newblock Framework model for packet loss metrics based on loss runlengths.
\newblock In {\em Electronic Imaging}, pages 177--187. International Society
  for Optics and Photonics, 1999.

\bibitem{savage2000practical}
Stefan Savage, David Wetherall, Anna Karlin, and Tom Anderson.
\newblock Practical network support for ip traceback.
\newblock In {\em ACM SIGCOMM Computer Communication Review}, volume~30, pages
  295--306. ACM, 2000.

\bibitem{schulz2014silence}
Steffen Schulz, Vijay Varadharajan, and Ahmad-Reza Sadeghi.
\newblock The silence of the lans: efficient leakage resilience for ipsec vpns.
\newblock {\em IEEE Transactions on Information Forensics and Security},
  9(2):221--232, 2014.

\bibitem{song2001advanced}
Dawn~Xiaodong Song and Adrian Perrig.
\newblock Advanced and authenticated marking schemes for ip traceback.
\newblock In {\em INFOCOM 2001. Twentieth Annual Joint Conference of the IEEE
  Computer and Communications Societies. Proceedings. IEEE}, volume~2, pages
  878--886. IEEE, 2001.

\bibitem{sung2003ip}
Minho Sung and Jun Xu.
\newblock Ip traceback-based intelligent packet filtering: a novel technique
  for defending against internet ddos attacks.
\newblock {\em IEEE Transactions on Parallel and Distributed Systems},
  14(9):861--872, 2003.

\bibitem{wang2010double}
Xiaogang Wang, Junzhou Luo, and Ming Yang.
\newblock A double interval centroid-based watermark for network flow
  traceback.
\newblock In {\em Computer Supported Cooperative Work in Design (CSCWD), 2010
  14th International Conference on}, pages 146--151. IEEE, 2010.

\bibitem{wang2007network}
Xinyuan Wang, Shiping Chen, and Sushil Jajodia.
\newblock Network flow watermarking attack on low-latency anonymous
  communication systems.
\newblock In {\em Security and Privacy, 2007. SP'07. IEEE Symposium on}, pages
  116--130. IEEE, 2007.

\bibitem{wang2003robust1}
Xinyuan Wang and Douglas~S Reeves.
\newblock Robust correlation of encrypted attack traffic through stepping
  stones by manipulation of interpacket delays.
\newblock In {\em Proceedings of the 10th ACM conference on Computer and
  communications security}, pages 20--29. ACM, 2003.

\bibitem{wang2001sleepy}
Xinyuan Wang, Douglas~S Reeves, S~Felix Wu, and Jim Yuill.
\newblock Sleepy watermark tracing: An active network-based intrusion response
  framework.
\newblock {\em Trusted Information}, pages 369--384, 2001.

\bibitem{yu2007dsss}
Wei Yu, Xinwen Fu, Steve Graham, Dong Xuan, and Wei Zhao.
\newblock Dsss-based flow marking technique for invisible traceback.
\newblock In {\em Security and Privacy, 2007. SP'07. IEEE Symposium on}, pages
  18--32. IEEE, 2007.

\bibitem{yu2005accuracy}
Xunqi Yu, James~W Modestino, and Xusheng Tian.
\newblock The accuracy of gilbert models in predicting packet-loss statistics
  for a single-multiplexer network model.
\newblock In {\em INFOCOM 2005. 24th Annual Joint Conference of the IEEE
  Computer and Communications Societies. Proceedings IEEE}, volume~4, pages
  2602--2612. IEEE, 2005.

\bibitem{zand2014rippler}
Ali Zand, Giovanni Vigna, Richard Kemmerer, and Christopher Kruegel.
\newblock Rippler: Delay injection for service dependency detection.
\newblock {\em INFOCOM, 2014 Proceedings IEEE}, pages 2157--2165, 2014.

\end{thebibliography}

\end{document}